\documentclass[preprint,3p]{elsarticle}
\usepackage{soul}
\usepackage{hyperref}
\usepackage{amsmath}
\usepackage{graphicx}
\usepackage{subfig}
\usepackage{array}
\usepackage{amsfonts}
\usepackage[dvipsnames]{xcolor}
\usepackage{booktabs}
 
\journal{} 

\begin{document}

\begin{frontmatter}

\title{
Axisymmetric Charge-Conservative Electromagnetic Particle Simulation Algorithm on
Unstructured Grids: \\ Application to Microwave Vacuum Electronic Devices
}

\author[ad1]{Dong-Yeop Na}
\ead{na.94@osu.edu}
\author[ad2]{Yuri A. Omelchenko}
\ead{omelche@gmail.com}
\author[ad3]{Haksu Moon}
\ead{haksu.moon@gmail.com}
\author[ad4]{Ben-Hur V. Borges}
\ead{benhur@sc.usp.br}
\author[ad1]{Fernando L. Teixeira}
\ead{teixeira.5@osu.edu}

\address[ad1]{ElectroScience Laboratory and Department of Electrical and Computer Engineering, The Ohio State University, Columbus, OH 43212, USA}
\address[ad2]{Trinum Research Inc., San Diego, CA 92126, USA}
\address[ad3]{Intel Corporation, Hillsboro, OR 97124, USA}
\address[ad4]{Department of Electrical and Computer Engineering, University of Sao Paulo, Sao Carlos, SP 13566-590, Brazil}

\begin{abstract}
We present a charge-conservative electromagnetic particle-in-cell (EM-PIC) algorithm optimized for the analysis of vacuum electronic devices (VEDs) with cylindrical symmetry (axisymmetry).
We exploit the axisymmetry present in the device geometry, fields, and sources to reduce the dimensionality of the problem from 3D to 2D. Further, we employ `transformation optics' principles to map the original problem in polar coordinates with metric tensor $\mathrm{diag}(1, \rho^2, 1)$ to an equivalent problem on a Cartesian metric tensor $\mathrm{diag}(1,1,1)$ with an effective (artificial) inhomogeneous medium introduced. The resulting problem in the meridian ($\rho z$) plane is discretized using an unstructured 2D mesh considering $\text{TE}^{\phi}$-polarized fields. 
Electromagnetic field and source (node-based charges and edge-based currents) variables are expressed as differential forms of various degrees, and discretized using Whitney forms.
Using leapfrog time integration, we obtain a mixed $\mathcal{E}-\mathcal{B}$ finite-element time-domain scheme for the full-discrete Maxwell's equations. We achieve a local and explicit time update for the field equations by employing the sparse approximate inverse (SPAI) algorithm. 
Interpolating field values to particles' positions for solving Newton-Lorentz equations of motion is also done via Whitney forms. 
Particles are advanced using the Boris algorithm with relativistic correction.
A recently introduced charge-conserving scatter scheme tailored for 2D unstructured grids is used in the scatter step. 
The algorithm is validated considering cylindrical cavity and space-charge-limited cylindrical diode problems.
We use the algorithm to investigate the physical performance of VEDs designed to harness particle bunching effects arising from the coherent (resonance) Cerenkov electron beam interactions within micro-machined slow wave structures.
\end{abstract}


\end{frontmatter}


\section{Introduction}\label{sec:introduction}
Historically the need for high-power electromagnetic (EM) radiation sources in the gigahertz and terahertz frequency ranges has triggered significant technical advances in vacuum electronic devices (VED) \cite{gold1997review,cairns1997generation,booske2011vacuum,booske2008plasma,li2013experimental}, such as the gyrotron, free electron Laser, and traveling wave tube (TWT). These devices serve as a basis for a variety of applications in radar and communications systems, plasma heating for fusion, and radio-frequency (RF) accelerators \cite{gaponov1994applications,schamiloglu2004high}.

\begin{figure}[t]
\centering
	\subfloat[\label{fig:bwo_system}]{\includegraphics[width=3in]{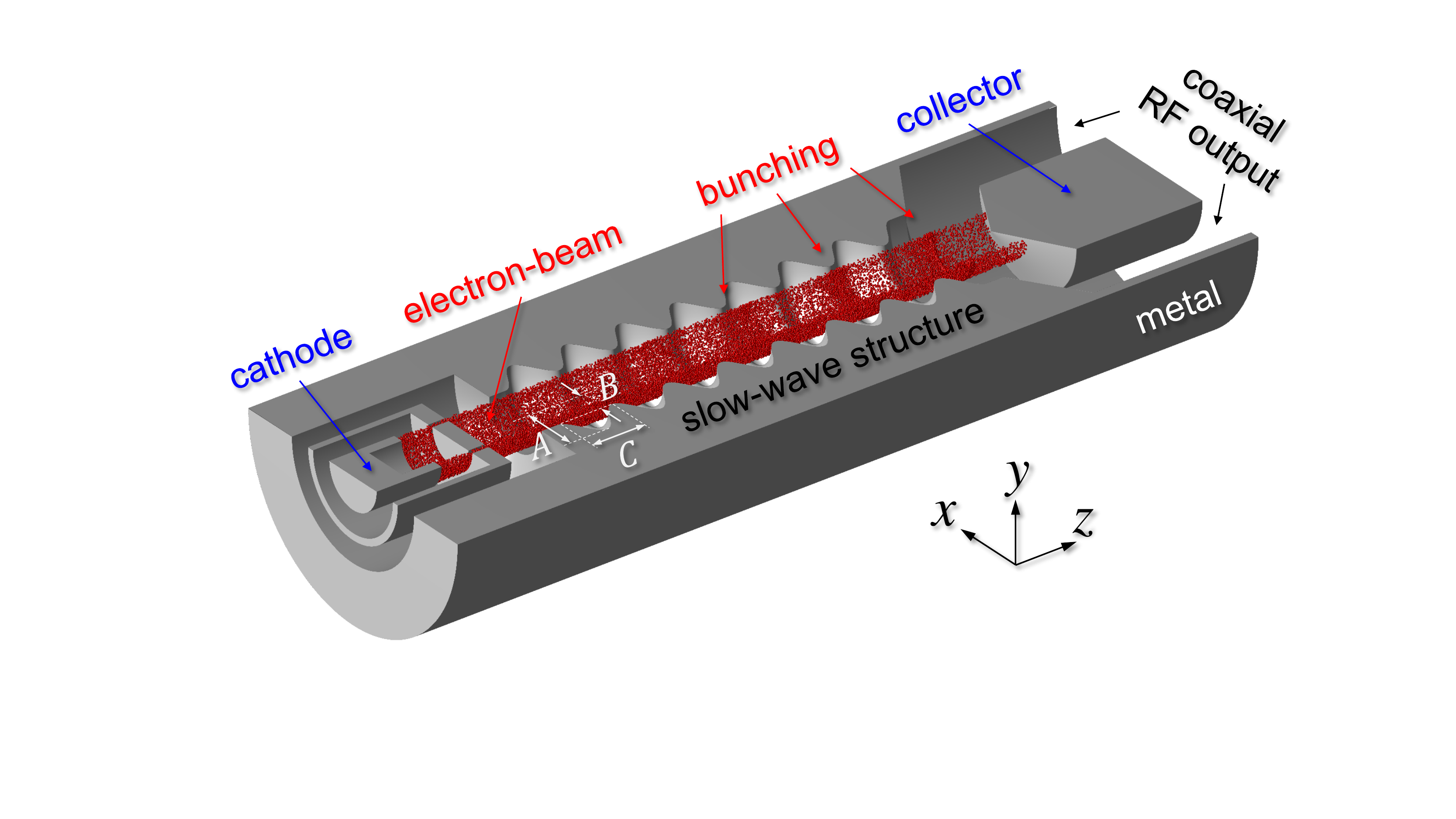}	
	}
	\subfloat[\label{fig:cyl_diode_system}]{\includegraphics[width=1.5in]{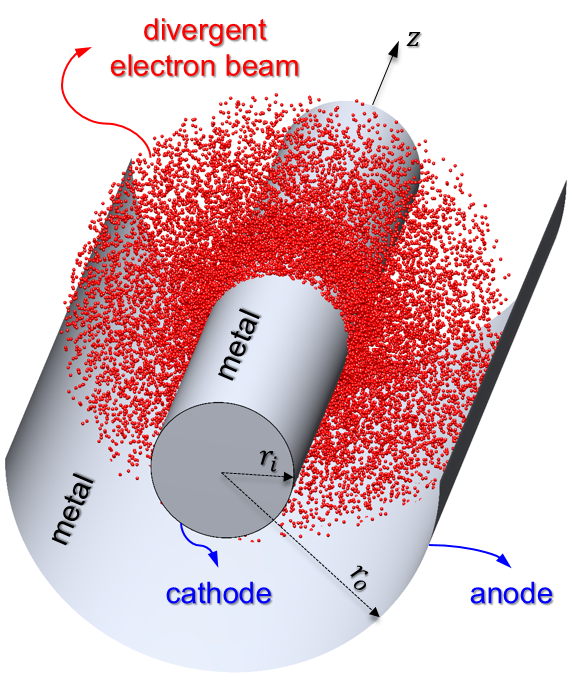}		
	}	
	\caption{Schematics of two examples of axisymmetric vacuum electronic devices. (a) Backward-wave oscillator producing bunching effects on an electron beam. Wall ripples are designed to support slow-wave modes in the device. (b) Space-charge-limited cylindrical vacuum diode.}
	\label{fig:objects}
\end{figure}

Amplification of RF signals is usually obtained by exploiting resonance Cerenkov interactions between an electron beam and the modal field supported by a slow-wave structure (SWS) \cite{cairns1997generation,shiffler1990high,johnson1955backward,gunin1998relativistic,case1984space}.
SWSs are often made by imposing periodic ripples on the conducting wall of cylindrically symmetric waveguides, as illustrated in Fig.~\ref{fig:bwo_system}, so that the phase velocity of the modal field becomes slower than the speed of light in vacuum due to Bragg scattering~\cite{chew1995waves}.
According to the dispersion relations associated with the geometry of SWSs, resultant Cerenkov interactions can amplify forward or backward waves\footnote{Along the direction of the group velocity w. r. t. the beam velocity.}.
Similarly to plasma instabilities, the evolution of forward and backward waves can be characterized by {\it convective} instabilities that grow over time while traveling away from the location of initial disturbance and {\it absolute} instabilities that propagate a local initial disturbance throughout the whole device volume~\cite{cairns1997generation}.
Traveling-wave tube amplifiers (TWTA) and backward-wave oscillators (BWO) are two practical examples utilizing convective and absolute instabilities, respectively.

Recent studies have shown that a particular SWS geometry may significantly enhance the system performance of TWTs.
For example, nonuniform (locally periodic) ripples used in BWOs may improve mode conversion efficiency \cite{moreland1994efficiency,chipengo2015novel}, and tapering ripples may reduce reflections at the output of TWTA and prevent internal oscillations \cite{shiffler1990high,shiffler1991high}. 
More importantly, smooth device edges are preferred for high-output power applications in order to mitigate pulse shortening, which is a major bottleneck for increasing output powers beyond the gigawatts range \cite{agee1998evolution,korovin2000pulsewidth}.
This is because extremely strong field singularities, which accumulate on the sharp edges, may create interfering plasmas that terminate the output signal at an earlier time.
Sinusoidally corrugated slow wave structures (SCSWS) have been increasingly adopted in many modern high-power BWO systems to combat this problem \cite{vlasov2000overmoded}. In addition, a variety of micro-machining fabrication techniques have been developed to enable better device performances by using much tighter tolerances. These technological advances have allowed the production of devices operating at higher frequencies, including the THz regime.

Computational experiments for VEDs employ electromagnetic particle-in-cell (EM-PIC) algorithms \cite{birdsall2004plasma,hockney1988computer,grigoryev2002numerical,candel2010parallel}, which numerically solve the Maxwell-Vlasov equations describing weakly coupled (collision-less) systems, where the collective behavior of charged particles prevails over their binary collisions \cite{birdsall2004plasma,hockney1988computer,dawson1983particle}.
A typical PIC algorithm tracks the temporal evolution of macro-particles seeded in a coarse-grained six-dimensional (6D) phase space\footnote{That is, a finite-size ensemble of physical particles with positions and momenta.}. A typical PIC algorithm consists of four basic steps, viz. the field solver, the field gather, the particle push, and the particle charge/current scatter, which are repeated at every time iteration. This provides a self-consistent update of particles and field states in time.

As a field solver, most EM-PIC simulations employ the celebrated Yee's finite-difference time-domain (FDTD) method for regular grids, due to its simplicity.
There are a plethora of FDTD-based EM-PIC codes such as \texttt{UNIPIC}, \texttt{MAGIC}, \texttt{TWOQUICK}, \texttt{KARAT}, \texttt{VORPAL}, and others \cite{wang2009unipic,wang2010three,nieter2009application}.
However, the relatively poor grid-dispersion properties of this algorithm \cite{taflove2000computational} causes spurious numerical Cerenkov radiation \cite{greenwood2004elimination}. Moreover, in complex geometries such as those of modern VEDs, ``staircase" (step-cell) effects present a critical challenge. Using FDTD for an accurate analysis of geometrically complex devices, which typically have curved boundaries or very fine geometrical features, may require excessive mesh refinement and therefore result in a waste of computational resources.
Many studies have been done to mitigate the staircasing errors in finite-difference (FD) methods, in particular through using conformal FD discretizations \cite{meierbachtol2015conformal,wang2016conformal}.

On the other hand, the finite-element time-domain (FETD) method~\cite{teixeira2008time,lee1997time} fundamentally eliminates the undesired staircase effects since 
it is naturally based on unstructured (irregular) grids, which can more easily be made conformal to complex geometries and can be augmented by powerful mesh refinement algorithms.
Unfortunately, conventional FETD-based PIC algorithms have historically faced numerical challenges that result from a lack of exact charge conservation on unstructured grids. 
This gives rise to the accumulation of spurious charges which must be removed by applying costly a posteriori corrections \cite{eastwood1991virtual,marder1987method}. 
In addition, implicit time updates used in conventional FETD require repeated linear solves at each time-step \cite{teixeira2008time}.
Recently, a novel charge-conserving scatter scheme for unstructured grids, inspired by differential-geometric ideas and the exterior calculus of differential forms \cite{burke1985, flanders1989,teixeira1999lattice,kotiuga2004,he2007differential,teixeira2014lattice}, has been proposed in \cite{moon2015exact}.
Other charge-conservative EM-PIC algorithms for unstructured grids were also developed under similar tenets in~\cite{squire2012geometric,pinto2014charge,pinto2016divergence}. In addition, a charge-conserving EM-PIC algorithm with explicit time-update that is both local (i.e. preserves sparsity) and obviates the need for linear solvers at each time step has been described in~\cite{he2006sparse,kim2011parallel,na2016local} based on the sparse approximate inverse (SPAI) of the discrete Hodge operator (the finite-element ``mass'' matrix).

These recent advances have made possible the present work, which is motivated by the demand to accurately capture realistic physics of beam-SWS interactions in complex geometry devices. In this paper we present a charge-conservative EM-PIC algorithm based on unstructured grids and optimized for the analysis and design of axisymmetric VEDs.
Since conventional SWSs are cylindrically axisymmetric (invariant along $\phi$), SWS studies can be best done with algorithms that explore this symmetry, so that significant computational resources can be saved and the algorithm may be feasibly implemented as a forward engine in a design loop~\cite{wang2016conformal}. We show that under the assumption of cylindrical symmetry of fields and sources one can reduce the original 3D geometry to a 2.5D setup by introducing an artificial inhomogeneous medium\footnote{In a manner akin to the ``transformation optics" (TO) technique~\cite{teixeira1999lattice,teixeira1999differential,pendry2006controlling,chen2010nature}.}, considering $\text{TE}^{\phi}$-polarized fields in the meridian ($\rho z$) plane. 
The spatial discretization of EM variables, which are represented by differential forms of various degrees, is achieved through applying discrete Whitney forms on an unstructured mesh in the meridian plane. 
Using leapfrog time integration, we obtain space- and time-discretized Maxwell's equations, a so-called mixed $\mathcal{E}-\mathcal{B}$ finite-element time-domain (FETD) scheme, for the field solver.
A local and explicit time-update requiring no repeated linear solver is made possible by employing a SPAI algorithm \cite{na2016local}. 
The interpolation of field values to particles' positions for solving the Newton-Lorentz equations of motion is also done via Whitney forms.
Relativistic particles are accelerated and pushed in space with a corrected Boris algorithm \cite{verboncoeur2005particle,vay2008simulation}. 
In the scatter step, we adapt an exact charge-conserving scheme developed for Cartesian unstructured grids \cite{moon2015exact}. 
We validate our algorithm using analytical results for a cylindrical cavity and previously obtained results for a space-charge-limited (SCL) vacuum diode (see Fig. \ref{fig:cyl_diode_system}).
We include a micro-machined SWS-based BWO example, designed to harness particle bunching effects from coherent Cerenkov beam-wave interaction,
 to demonstrate the advantages of utilizing unstructured grids without staircasing error to predict the device performance.

{\color{black}
\section{Spatial dimensionality reduction} \label{sec:reduction_dimension}
In this section we describe a numerical model for 3D VEDs with cylindrical axisymmetry based on an equivalent 2D model discretized on an unstructured (irregular) grid in the meridian plane.

\subsection{Exterior calculus representation of Maxwell's equations}\label{sec:Maxwell's_equation}
We represent Maxwell's equations using the exterior calculus of differential forms~\cite{flanders1989,kotiuga2004,teixeira2014lattice,squire2012geometric,FEMSTER2005} as
\begin{eqnarray}\
d\mathcal{E}\!\!\!\!&=&\!\!\!\!-\frac{\partial \mathcal{B}}{\partial t},
\\
d\mathcal{H}\!\!\!\!&=&\!\!\!\!\frac{\partial \mathcal{D}}{\partial t}+\mathcal{J},
\\
d\mathcal{D}\!\!\!\!&=&\!\!\!\!\mathcal{Q},
\\
d\mathcal{B}\!\!\!\!&=&\!\!\!\!0,
\end{eqnarray}
where $\mathcal{E}$ and $\mathcal{H}$ are 1-forms for the electric and magnetic field intensity, $\mathcal{D}$ and $\mathcal{B}$ are 2-forms for the electric and magnetic flux density, $\mathcal{J}$ is 2-form for the electric current density, $\mathcal{Q}$ is 3-form for the electric charge density, and operator $d$ is the exterior derivative encompassing conventional curl and divergence operators~\cite{warnick1997teaching,sen2000geometric,bossavit1988whitney,teixeira2013differential}.
These 1-, 2-, and 3-forms can be expressed using a set of non-orthonormal-basis in a cylindrical coordinate system $\left(d\rho,d\phi,dz\right)$ \cite{teixeira1999differential,warnick1997teaching}.
For instance, $\mathcal{E}$ (1-form) is expressed as $\mathcal{E}=E_{\rho}d\rho+E_{\phi}d\phi+E_{z}dz$; then, its vector proxy can be written as $\vec{E}=E_{\rho}\hat{\rho}+\frac{E_{\phi}}{\rho}\hat{\phi}+E_{z}\hat{z}$.  
Similarly, $\mathcal{B}$ (2-form) is given by $\mathcal{B}=B_{\rho}d\phi\wedge dz+B_{\phi}dz\wedge d\rho+B_{z}d\rho\wedge d\phi$ (where $\wedge$ is the exterior or wedge product~\cite{burke1985, flanders1989,sen2000geometric}) and its vector proxy takes the form of $\vec{B}=\frac{B_{\rho}}{\rho}\hat{\rho}+B_{\phi}\hat{\phi}+\frac{B_{z}}{\rho}\hat{z}$~\cite{warnick1997teaching}. 

\subsection{Cylindrical axisymmetry constraints}\label{sec:Cylindrical axisymmetry constraints}
Consider a charged ring with constant density along azimuth that travels inside of a cylindrically axisymmetric tube bounded by a perfect electric conductor (PEC) with a radial boundary profile of $\partial \Omega=R\left(z\right)$ where $\Omega$ is the computational domain and $R$ is the wall radius which only depends on $z$, as shown in Fig.~\ref{fig:geom_a}.
\begin{figure}[t]
    \centering
	\subfloat[\label{fig:geom_a}]{
     \includegraphics[width=3.5in]{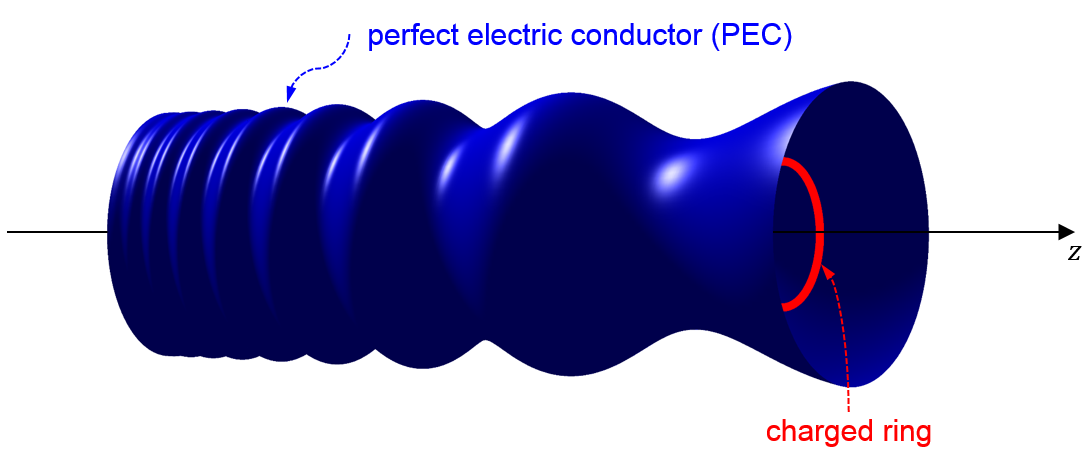}	
	}
\\
	\subfloat[\label{fig:geom_b}]{
     \includegraphics[width=3.5in]{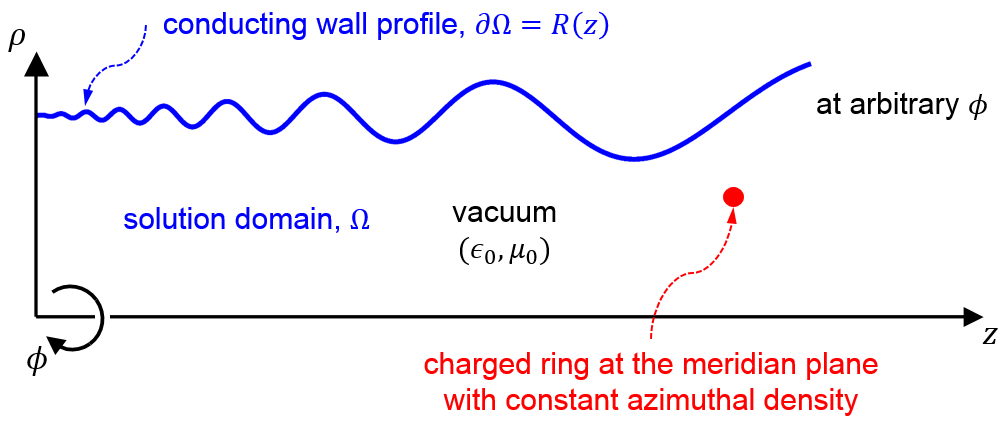}
	}		
    \caption{A charged ring travels inside an axisymmetric object bounded by PEC: (a) a 3D view, (b) the meridian plane.}
    \label{fig:geom}
\end{figure}
{\it Cylindrical axisymmetry}, used here, implies that there is no variation along $\phi$ (${\partial}/{\partial \phi}=0$) in the device geometry, fields, and sources. It should be noted that axisymmetric sources in the meridian plane are represented as charge rings (see Fig. \ref{fig:geom_b}). There exist two useful constraints that simplify the original 3D problem: (i) elimination of the $d\phi\frac{\partial}{\partial \phi}$ term in the exterior derivative $d$, viz. $d=d\rho\frac{\partial}{\partial \rho}+dz\frac{\partial}{\partial z}$ and (ii) retainment of only transverse magnetic (TM) eigenmodes with $m=0$\footnote{The index $m$ is used here to denote azimuthal harmonics.}.
The first constraint enables the same calculus in the meridian plane as in the 2D Cartesian coordinate system with the cylindrical metric factor embedded into the constitutive relations, as discussed below. 
The second constraint simplifies expressions for fields and sources as
\begin{eqnarray}
\label{eq:E_expr}
\mathcal{E}\!\!\!\!&=&\!\!\!\!E_{\rho}d\rho+E_{z}dz,
\\
\label{eq:B_expr}
\mathcal{B}\!\!\!\!&=&\!\!\!\!B_{\phi}dz\wedge d\rho,
\\
\label{eq:D_expr}
\mathcal{D}\!\!\!\!&=&\!\!\!\!D_{\rho}d\phi\wedge dz+D_{z}d\rho\wedge d\phi,
\\
\label{eq:H_expr}
\mathcal{H}\!\!\!\!&=&\!\!\!\!H_{\phi} d\phi.
\end{eqnarray}
From equations of (\ref{eq:E_expr}), (\ref{eq:B_expr}), (\ref{eq:D_expr}), (\ref{eq:H_expr}), it is straightforward to show that a 3D problem with cylindrical axisymmetry can be represented by a 2D problem describing $\text{TE}^{\phi}$-polarized fields on the meridian plane (see also Fig. \ref{fig:equiv_prob}).

\subsection{Modified Hodge star operator}\label{sec:Modification of the Hodge star operator}
The Hodge star operator $\star$  maps $p$-forms into $\left(n-p\right)$-forms in $n$-dimensional space\footnote{The Hodge star operator can be understood geometrically as yielding the orthogonal complement of a given differential form
to the volume form in $n$-space. In the 3D case for example,   
 $\star (dx)$ = $dy \wedge dz$,  $ \star (dz \wedge d\rho)$ = $\rho d \phi$, $\star (1)$ = $dx \wedge dy \wedge dz$ = $\rho \, d\rho \wedge d\phi \wedge dz$ and so forth~\cite{warnick1997teaching}. For an arbitrary $p$-form 
 ${\cal A}$ in $n$-space, we have ${\cal A} \wedge (\star {\cal A}) = |A|^2 dV$, where $dV$ is the volume element in $n$-space. 
 As such,  ${\cal A} \wedge (\star {\cal A})$ provides the $L^2$ ( or``energy'') element norm of  ${\cal A}$.}~\cite{he2007differential,teixeira2014lattice,warnick1997teaching,teixeira2013differential,tarhasaari1999some,gillette2011dual}.  In our case, we have
\begin{eqnarray}
\label{eq:HS_H}
\mathcal{H}\!\!\!\!&=&\!\!\!\!{\mu_{0}^{-1}}\star\mathcal{B},
\\
\label{eq:HS_D}
\mathcal{D}\!\!\!\!&=&\!\!\!\!{\epsilon_{0}}\star\mathcal{E}.
\end{eqnarray}
The Hodge operators incorporate the metrical properties of the system, which in the cylindrical case are 
expressed by a metric tensor $\mathrm{diag}(1,\rho^2,1)$.
For the magnetic field and flux density, substituting (\ref{eq:H_expr}) into the left-hand side term of (\ref{eq:HS_H}) gives
\begin{eqnarray}
\label{eq:HS_H_LHS}
\mathcal{H}=H_{\phi} d\phi,
\end{eqnarray}
and substituting (\ref{eq:B_expr}) into the right-hand side term of (\ref{eq:HS_H}) yields
\begin{eqnarray}
\label{eq:HS_H_RHS}
\!\!\!\!\!\!\!\!\!\!\!\!
{\mu_{0}^{-1}}\star\mathcal{B}
\!\!\!\!&=&\!\!\!\!
\mu_{0}^{-1}\star\left(B_{\phi}dz\wedge d\rho\right)
=
\mu_{0}^{-1}B_{\phi}\star\left(dz\wedge d\rho\right)
\nonumber \\
\!\!\!\!&=&\!\!\!\!
\mu_{0}^{-1}B_{\phi}\rho d\phi
=
\left(\mu_{0}^{-1}\rho\right)B_{\phi}d\phi
=
\mu^{-1}\left(\rho\right)B_{\phi}d\phi.
\end{eqnarray}
Note that $\star$ only acts on the differentials such as $dz$, $d\rho$, and $d\phi$.
By comparing (\ref{eq:HS_H_LHS}) and (\ref{eq:HS_H_RHS}) and introducing an artificial magnetic permeability, $\mu\left(\rho\right)=\mu_{0}\rho^{-1}$, we can extract the radial factor $\rho$ from the Hodge star operator. As a result, we can reuse simple Cartesian space calculus with metric tensor $\mathrm{diag}(1,1,1)$ and a Hodge operator devoid of additional metric factors.
For the electric field and flux density, substituting (\ref{eq:D_expr}) into the left-hand side term of (\ref{eq:HS_D}) yields
\begin{eqnarray}
\!\!\!\!\!\!\!\!\!\!\!\!
\mathcal{D}=D_{\rho} d\phi\wedge dz\!\!\!\!&+&\!\!\!\!D_{z}d\rho\wedge d\phi,
\end{eqnarray}
and substituting (\ref{eq:E_expr}) into the right-hand side term of (\ref{eq:HS_D}) gives
\begin{eqnarray}
\label{eq:HS_D_RHS}
\!\!\!\!\!\!\!\!\!\!\!\!
{\epsilon_{0}}\star\mathcal{E}
\!\!\!\!&=&\!\!\!\!
\epsilon_{0}\star\left(E_{\rho}d\rho+E_{z}dz\right)
=
\epsilon_{0}\left(E_{\rho}\star d\rho+E_{z}\star dz\right)
\nonumber\\
\!\!\!\!&=&\!\!\!\! \epsilon_{0}\left(E_{\rho} \rho d\phi\wedge dz+E_{z}  d\rho\wedge \rho d\phi\right)
=
\left(\epsilon_{0}\rho\right)\left(E_{\rho} d\phi\wedge dz+E_{z} d\rho\wedge d\phi\right)
\nonumber\\
\!\!\!\!&=&\!\!\!\!\epsilon\left(\rho\right)\left(E_{\rho} d\phi\wedge dz+E_{z} d\rho\wedge d\phi\right).
\end{eqnarray}
In a similar fashion, an artificial electrical permittivity takes the form of $\epsilon\left(\rho\right)=\epsilon_{0}\rho$.
Essentially, an original 3D problem with cylindrical axisymmetry is replaced by an equivalent 2D problem with $\text{TE}^{\phi}$-polarized fields immersed in Cartesian space with an inhomogeneous medium with artificial permittivity and permeability, as illustrated schematically in Fig. \ref{fig:equiv_prob}.
\begin{figure}[t]
    \centering
	\includegraphics[width=3.5in]{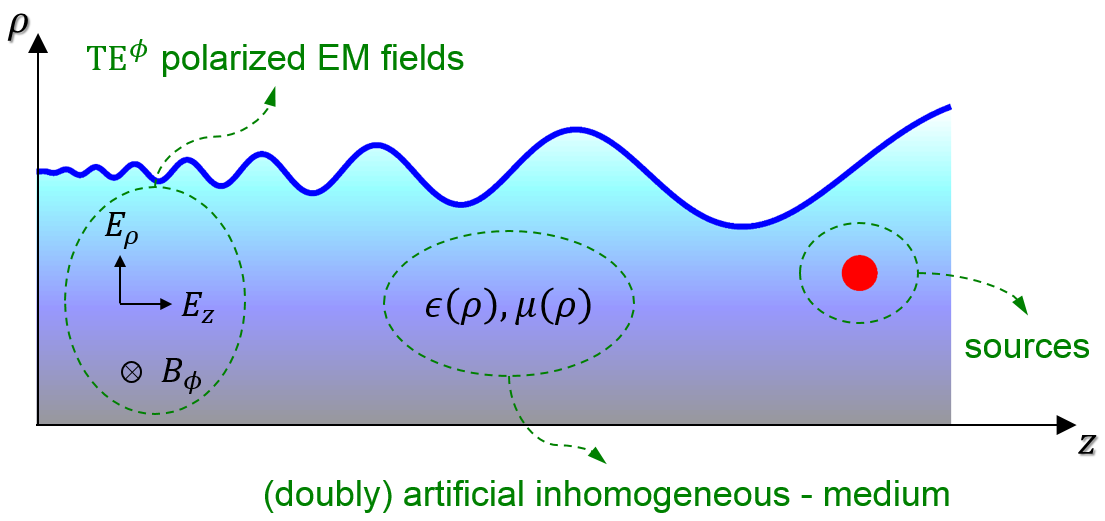}
    \caption{The original problem shown in Fig. \ref{fig:geom} is replaced by an equivalent 2D problem in the meridian plane as depicted above, which considers $\text{TE}^{\phi}$-polarized EM fields on Cartesian space with an artificial inhomogeneous medium. The variable coloring serves to stress the dependency of the artificial medium parameters on $\rho$.}
    \label{fig:equiv_prob}
\end{figure}
As a result, the present field-solver borrowing the concept of TO is free from the axial singularity associated with the $1/\rho$ factor present in the cylindrical nabla operator.

{\color{black}
\section{An axisymmetric EM-PIC algorithm}
EM-PIC algorithms primarily operate in time domain. Their computational cycle typically consists of the following steps: (1) EM field update (solution of Maxwell's equations), (2) field gather (interpolating field values to the particle positions), (3) particle update (solution of the Newton-Lorentz equations of motion), and (4) particle scatter (computing grid currents and charges from the kinetic information stored in charged particles). 

\subsection{Field solver: Mixed $\mathcal{E}-\mathcal{B}$ FETD with local and explicit time update}\label{sec:FETD}
In the field update, space- and time-discretized Maxwell's equations are solved for all discrete EM fields in one time-step.
For the 2D equivalent problem discussed in Sec. \ref{sec:reduction_dimension}, we can define $\mathcal{E}$, $\mathcal{B}$, $\mathcal{D}$, and $\mathcal{H}$ as a 1-, 2-, 1-, and 0-forms, respectively, and $\mathcal{J}_{\star}$ and $\mathcal{Q}_{\star}$ as 1- and 0-forms~\footnote{Since $\mathcal{J}$ and $\mathcal{Q}$ are 1- and 2-forms, respectively.}.
It should be stressed that the operator $\star$ used here for the equivalent problem is defined in the Cartesian 2D space and hence devoid of the radial factor $\rho$ (which enters instead in the definition of the artificial medium properties). 
Because $\mathcal{E}$ and $\mathcal{B}$ are ordinary differential forms with internal orientations associated with the primal lattice~\cite{teixeira2014lattice,teixeira2013differential}, we expand them by using Whitney 1- and 2-forms (cochains) residing in the primal mesh, as follows:
\begin{eqnarray}
\mathcal{E}(t)=\sum_{i=1}^{N_{1}} \mathbb{E}_{i}(t) \, {w}_{i}^{1},
\\
\mathcal{B}(t)=\sum_{i=1}^{N_{2}}\mathbb{B}_{i}(t) \, {w}_{i}^{2},
\end{eqnarray}
where $N_{p}$ is the number of $p$-cells for $p=0,1,2$ (number of nodes, edges, facets, respectively, on the grid), $w_{i}^{p}$ is the Whitney $p$-form associated with the $i$-th $p$-cell on the grid, and $\mathbb{E}_{i}(t)$ and $\mathbb{B}_{i}(t)$ are the (time-evolving) degrees of freedoms (DoFs) for $\mathcal{E}$ and $\mathcal{B}$, respectively~\cite{white2015mixed}.
On the other hand, $\mathcal{D}$, $\mathcal{H}$, $\mathcal{J}$, and $\mathcal{Q}$ are twisted differential forms with external orientations, associated with the dual lattice~\cite{teixeira1999lattice,teixeira2014lattice,sen2000geometric}. 
Fig. \ref{fig:pd_mesh} shows schematically how the DoFs are associated to the primal and dual meshes elements in our 2D setting.
\begin{figure}[t]
     \centering
	\subfloat[\label{fig:primal_mesh_qt}]{
     \includegraphics[width=2.25in]{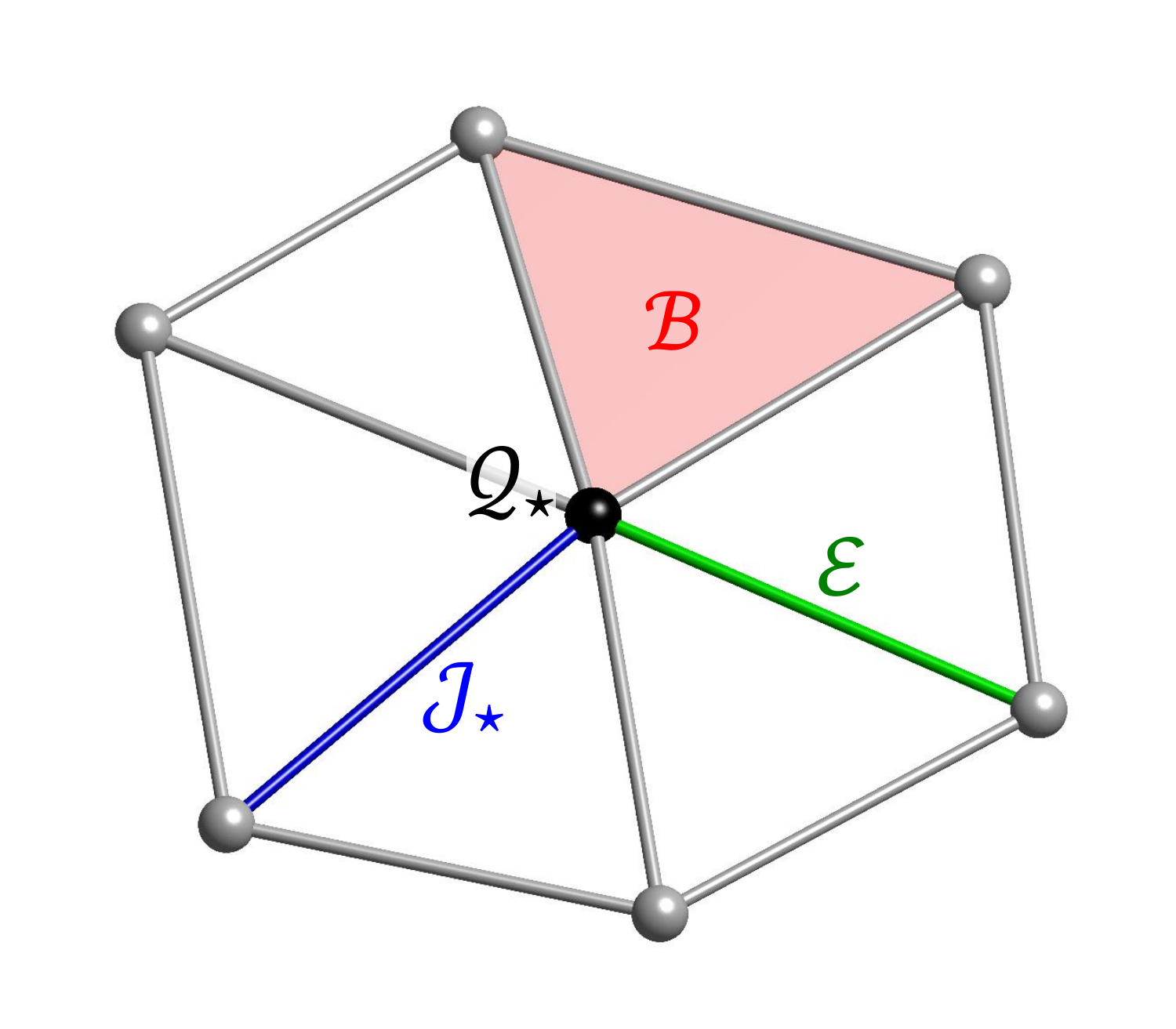}
     }
	\subfloat[\label{fig:dual_mesh_qt}]{
     \includegraphics[width=2.25in]{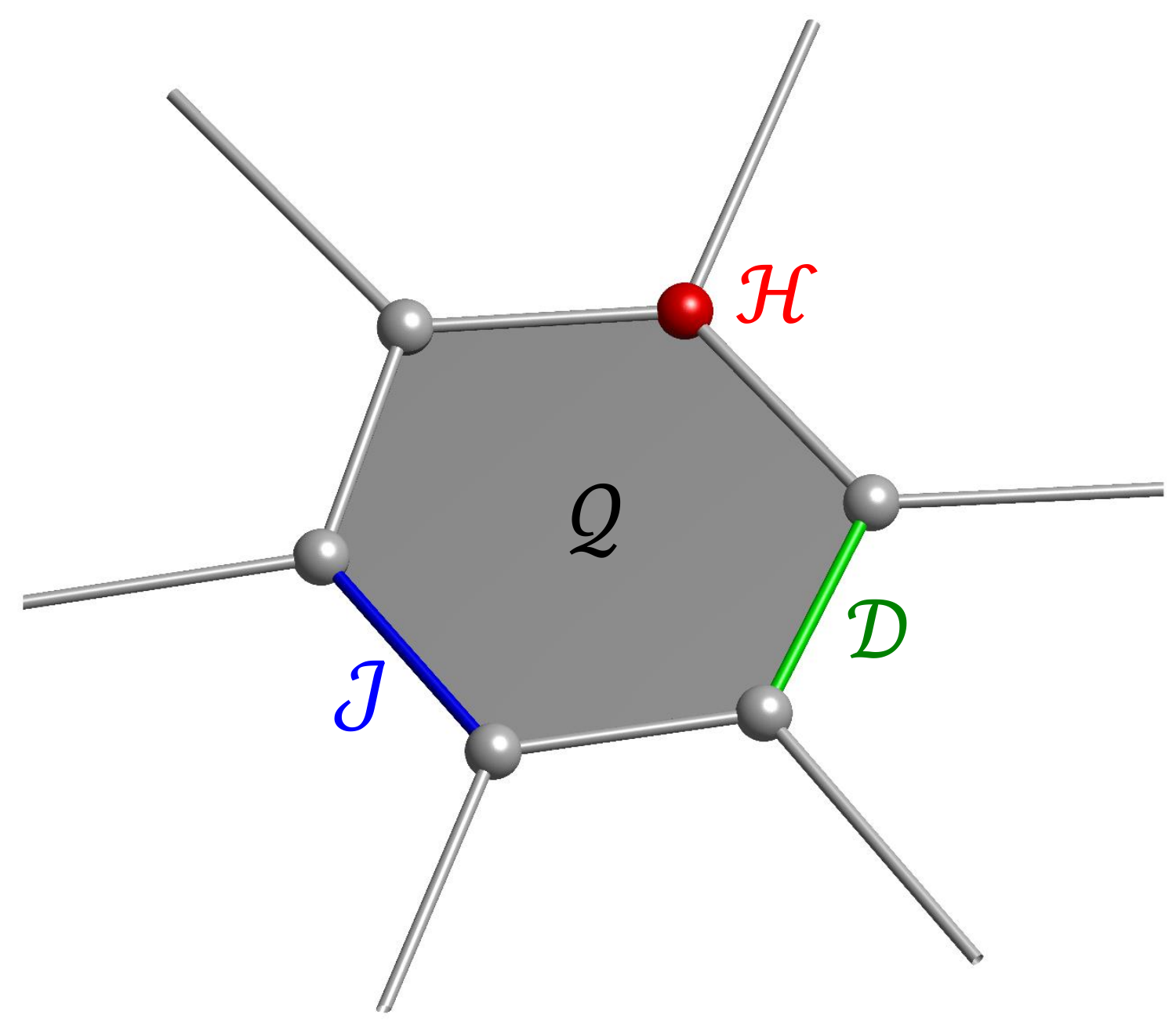}
     }
    \caption{The (2+1) setup for fields and sources defined in (a) primal and (b) dual meshes.}
    \label{fig:pd_mesh}
\end{figure}
To avoid the need for working with the dual mesh, we employ the Hodge duals ${\mathcal{J}}_{\star}$ and $\mathcal{Q}_{\star}$ which can be expanded using Whitney 1- and 0-forms in the primal mesh, as follows:
\begin{eqnarray}
\mathcal{J}_{\star}(t)=\sum_{i=1}^{N_{1}}\mathbb{J}_{\star,i}(t) \, w_{i}^{1},
\\
\mathcal{Q}_{\star}(t) =\sum_{i=1}^{N_{0}}\mathbb{Q}_{\star,i}(t) \, w_{i}^{0}
\end{eqnarray}
where $\mathbb{J}_{\star,i}$ and $\mathbb{Q}_{\star,i}$ are Dos of $i$-th edge and node for ${\mathcal{J}}_{\star}$ and $\mathcal{Q}_{\star}$, respectively.
Applying a pairing operation and generalized Stokes' theorem to Faraday's law in the primal mesh and Ampere's law in the dual mesh \cite{teixeira1999lattice,teixeira2014lattice,he2005degrees}, the spatially discretized Maxwell's equations (see also \ref{sec:app_A}) are given by
\begin{eqnarray}
\left[\mathcal{D}_{\text{curl}}\right]\cdot\left[\mathbb{E}\right]
\!\!\!\!&=&\!\!\!\!
-\frac{\partial}{\partial t}\left[\mathbb{B}\right]
\label{eq:DFL_mat}
\\
\left[\mathcal{\tilde{D}}_{\text{curl}}\right]\cdot\left[\star_{\mu^{-1}}\right]\cdot\left[\mathbb{B}\right]
\!\!\!\!&=&\!\!\!\!
\frac{\partial}{\partial t}\left[\star_{\epsilon}\right]\cdot\left[\mathbb{E}\right]+\left[\mathbb{J}_{\star}\right],
\label{eq:DAL_mat}
\end{eqnarray}
with $\left[\mathbb{E}\right]=\left\{\mathbb{E}_{1},...,\mathbb{E}_{N_{1}}\right\}^{T}$, $\left[\mathbb{B}\right]=\left\{\mathbb{B}_{1},...,\mathbb{B}_{N_{2}}\right\}^{T}$, $\left[\mathbb{J}_{\star}\right]=\left\{\mathbb{J}_{\star,1},...,\mathbb{J}_{\star,N_{1}}\right\}^{T}$, and where $\left[\mathcal{D}_{\text{curl}}\right]$ and $\left[\mathcal{\tilde{D}}_{\text{curl}}\right]$  are incidence matrices encoding the discrete representation of the exterior derivative (or, equivalently, the curl operator distilled from the metric) on the primal and dual meshes, respectively~\cite{teixeira1999lattice,teixeira2014lattice,hughes1981lagrangian,guth1980existence}. 
Rows of $\left[\mathcal{D}_{\text{curl}}\right]$ have a $1:1$ association to faces of the primal mesh and columns to edges. On a primal triangular mesh, each row of
$\left[\mathcal{D}_{\text{curl}}\right]$ has three nonzero elements corresponding to the three edges of the triangular cell associated to that row (with entries $1$ or $-1$ depending on their relative orientation). All the other elements of that row (corresponding to edges that are not part of the boundary of the triangular cell boundary) are zero. A similar relationship exists for $\left[\mathcal{\tilde{D}}_{\text{curl}}\right]$  so that both of these matrices take their elements in the set of $\left\{-1,0,1\right\}$ as a consequence of their metric-free character. It can be shown that
 $\left[\mathcal{\tilde{D}}_{\text{curl}}\right]$=$\left[\mathcal{D}_{\text{curl}}\right]^T$~\cite{teixeira1999lattice}.
In (\ref{eq:DFL_mat}) and 
(\ref{eq:DAL_mat}),
$\left[\star_{\epsilon}\right]$ and $\left[\star_{\mu^{-1}}\right]$ are discrete Hodge operators (see \ref{sec:app_B}), which are symmetric positive-definite matrices obtained by the Galerkin method~\cite{kim2011parallel,tarhasaari1999some,gillette2011dual,he2006geometric}. 
Their elements are expressed as the area (volume in 3D) integrals
\begin{eqnarray}
\left[\star_{\epsilon}\right]_{i,j}
\!\!\!\!&=&\!\!\!\!
\int_{K}{\epsilon}w_{i}^{1}\wedge\star w_{j}^{1},
=
{\epsilon_{0}}\int_{K}\rho w_{i}^{1}\wedge\star w_{j}^{1},
\label{eq:dis_hodge_eps}
\\
\left[\star_{\mu^{-1}}\right]_{i,j}
\!\!\!\!&=&\!\!\!\!
\int_{K}{\mu^{-1}}w_{i}^{2}\wedge\star w_{j}^{2},
=
{\mu_{0}^{-1}}
\int_{K}\rho w_{i}^{2}\wedge\star w_{j}^{2},
\label{eq:dis_hodge_mu}
\end{eqnarray}
where $K$ is the intersection of the Whitney functions spatial support, respectively.
Vector proxies of (\ref{eq:dis_hodge_eps}) and (\ref{eq:dis_hodge_mu}) are provided in \ref{sec:app_B}. 
Note that the discrete Hodge matrix for the electric current density becomes the identity matrix for the barycentric dual mesh (see \ref{sec:app_C}).
We apply leapfrog time integration for time discretization with the electric field, flux density and charge density defined at integer time-steps $n$ and the magnetic flux density, field and current density defined at half integer time-steps $n\pm\frac{1}{2}$.
The resulting space- and time-discretized Maxwell's equations form the so-called mixed $\mathcal{E}-\mathcal{B}$ finite-element time-domain (FETD) scheme \cite{moon2015exact,kim2011parallel,na2016local,donderici2008mixed}:
\begin{eqnarray}
\!\!\!\!\!\!\!\!
\left[\mathbb{B}\right]^{n+\frac{1}{2}}
\!\!\!\!&=&\!\!\!\!
\left[\mathbb{B}\right]^{n-\frac{1}{2}}-\Delta t \left[\mathcal{D}_{\text{curl}}\right]\cdot\left[\mathbb{E}\right]^{n}
\label{eq:full_d_dfl}
\\
\!\!\!\!\!\!\!\!
\left[\mathbb{E}\right]^{n+1}
\!\!\!\!&=&\!\!\!\!
\left[\mathbb{E}\right]^{n}
+
\Delta t\left[\star_{\epsilon}\right]^{-1}\cdot
\left(
\left[\mathcal{\tilde{D}}_{\text{curl}}\right]\cdot\left[\star_{\mu^{-1}}\right]\cdot\left[\mathbb{B}\right]^{n+\frac{1}{2}}
-\left[\mathbb{J}_{\star}\right]^{n+\frac{1}{2}}\right),
\label{eq:full_d_dal}
\end{eqnarray}
where superscripts $n$ denote time-step and $\Delta t$ is the time-step increment.
Local explicit time-update is achieved using the SPAI approach to obtain a representation for $\left[\star_{\epsilon}\right]^{-1}$ and obviate the need for a (expensive) linear-solver at each time-step~\cite{kim2011parallel,na2016local}.
Since the discrete equations (\ref{eq:full_d_dfl}) and (\ref{eq:full_d_dal}) are basically sparse linear systems, matrix multiplications with compressed sparse row format are adopted to reduce computational costs.

\subsection{Field gather}\label{sec:Field gather}
The EM fields must be determined at the positions of $\alpha$-species particles in order to solve the (Newton-Lorentz) equations of motion for the latter. 
This is again done through using Whitney forms, which recover the necessary field values from their grid-defined representations as~\cite{moon2015exact}
\begin{eqnarray}
\!\!\!\!\!\!\!\!
\vec{E}^{\text{EM}}\left(\vec{r}_{\alpha},n\Delta t\right)
\!\!\!\!&=&\!\!\!\!
\vec{E}_{\alpha}^{\text{EM},n}
=
\sum_{i=1}^{N_{1}}\mathbb{E}^{n}_{i}\vec{W}_{i}^{1}\left(\vec{r}_{\alpha}\right),
\\
\!\!\!\!\!\!\!\!
\vec{B}^{\text{EM}}\left(\vec{r}_{\alpha},n\Delta t\right)=\vec{B}_{\alpha}^{\text{EM},n}
\!\!\!\!&=&\!\!\!\!
\sum_{i=1}^{N_{2}}
\left(
\frac{\mathbb{B}^{n+\frac{1}{2}}_{i}+\mathbb{B}^{n-\frac{1}{2}}_{i}}{2}
\right)
\vec{W}_{i}^{2}\left(\vec{r}_{\alpha}\right).
\end{eqnarray}
where $\vec{W}_{i}^{p}$ is a vector proxy of the Whitney $p$-form for $i$-th grid element and $\vec{r}_{\alpha}$ is the particle position vector.

A key component in the operation of BWOs is the beam focusing system (BFS) which axially confines an electron beam within an extremely small gyroradius in the polar ($\rho\phi$) plane to avoid the bombardment of the SWS by the beam electrons.
Beam focusing is usually achieved by applying a strong static axial magnetic field in the SWS region.
Because a strong confinement of the electron beam, {\it azimuthal} particle bunching effects can be ignored.

\subsection{Relativistic particle pusher}\label{sec:Relativistic particle pusher}
The positions and velocities of $\alpha$-species particles are updated by solving the Newton-Lorentz equations of motion as~\cite{na2016local}.
\begin{eqnarray}
\frac{d\vec{r}_{\alpha}}{dt}
\!\!\!\!&=&\!\!\!\!
\vec{v}_{\alpha},
\label{eq:NE}
\\
\frac{d\left(m_{\alpha}\gamma_{\alpha}\vec{v}_{\alpha}\right)}{dt}
\!\!\!\!&=&\!\!\!\!
q_{\alpha}\left[\vec{E}+\vec{v}_{\alpha}\times\vec{B}\right],
\label{eq:LE}
\end{eqnarray}
where $\vec{v}$ is a velocity vector, $m_{\alpha}$ is a particle mass, $q_{\alpha}$ is a particle charge, and $\gamma=\frac{1}{\sqrt{1-\left(\frac{\left|\vec{v}\right|}{c}\right)^{2}}}$, and $\vec{u}=\gamma\vec{v}$. 
The leapfrog time-discretized Newton-Lorentz equations of motion can be written as
\begin{eqnarray}
\frac{\vec{r}_{\alpha}^{n+1}-\vec{r}_{\alpha}^{n}}{\Delta_{t}}
\!\!\!\!&=&\!\!\!\!
\vec{v}_{\alpha}^{n+\frac{1}{2}},
\label{eq:DNE}
\\
\frac{\vec{u}_{\alpha}^{n+\frac{1}{2}}-\vec{u}_{\alpha}^{n-\frac{1}{2}}}{\Delta t}
\!\!\!\!&=&\!\!\!\!
\frac{q_{\alpha}}{m_{\alpha}}\left(\vec{E}_{\alpha}^{n}+\frac{\vec{u}_{\alpha}^{n}}{\gamma_{\alpha}^{n}}\times\vec{B}_{\alpha}^{n}\right).
\label{eq:DLE}
\end{eqnarray}
Here, the position and velocity vectors of particles are defined at integer and half-integer time-steps, respectively.
In order to solve the implicit equation (\ref{eq:DLE}), we adopt the Boris algorithm with a correction \cite{verboncoeur2005particle,vay2008simulation}.



\begin{figure}[t]
    \centering
	\subfloat[\label{fig:q_def_a}]{
    \includegraphics[width=1.1in]{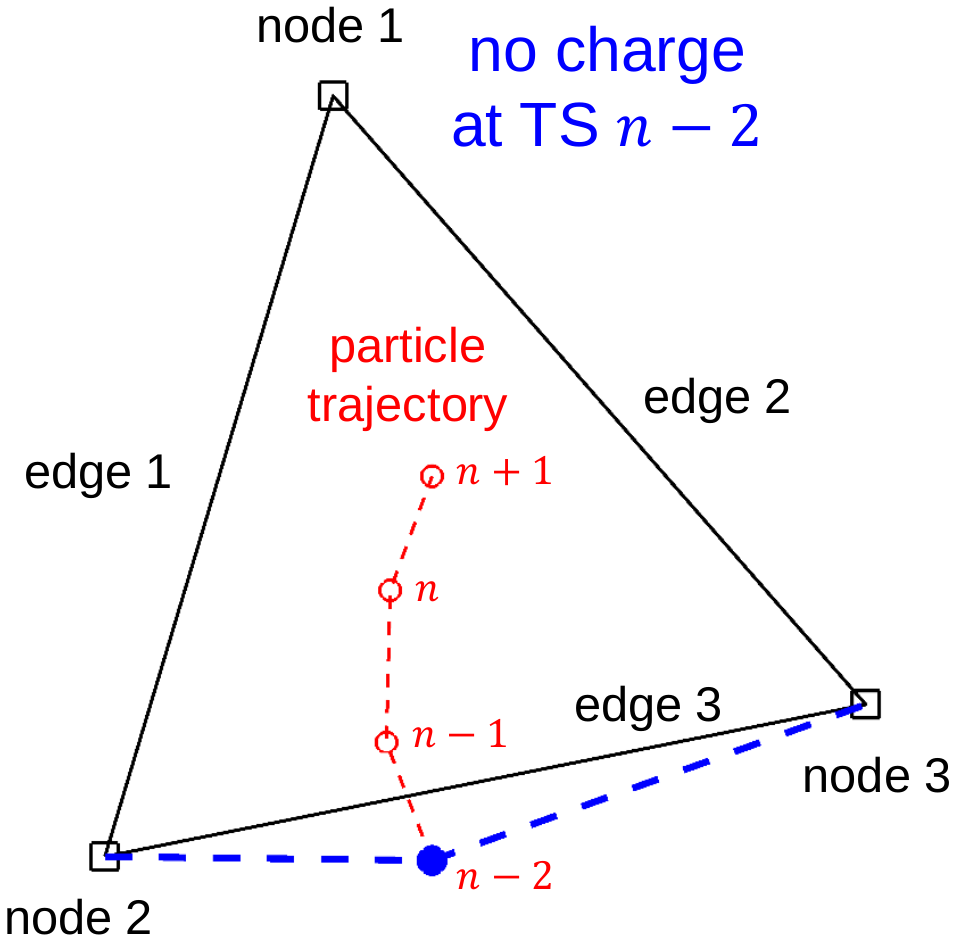}
    }
	\subfloat[\label{fig:q_def_b}]{
    \includegraphics[width=1.1in]{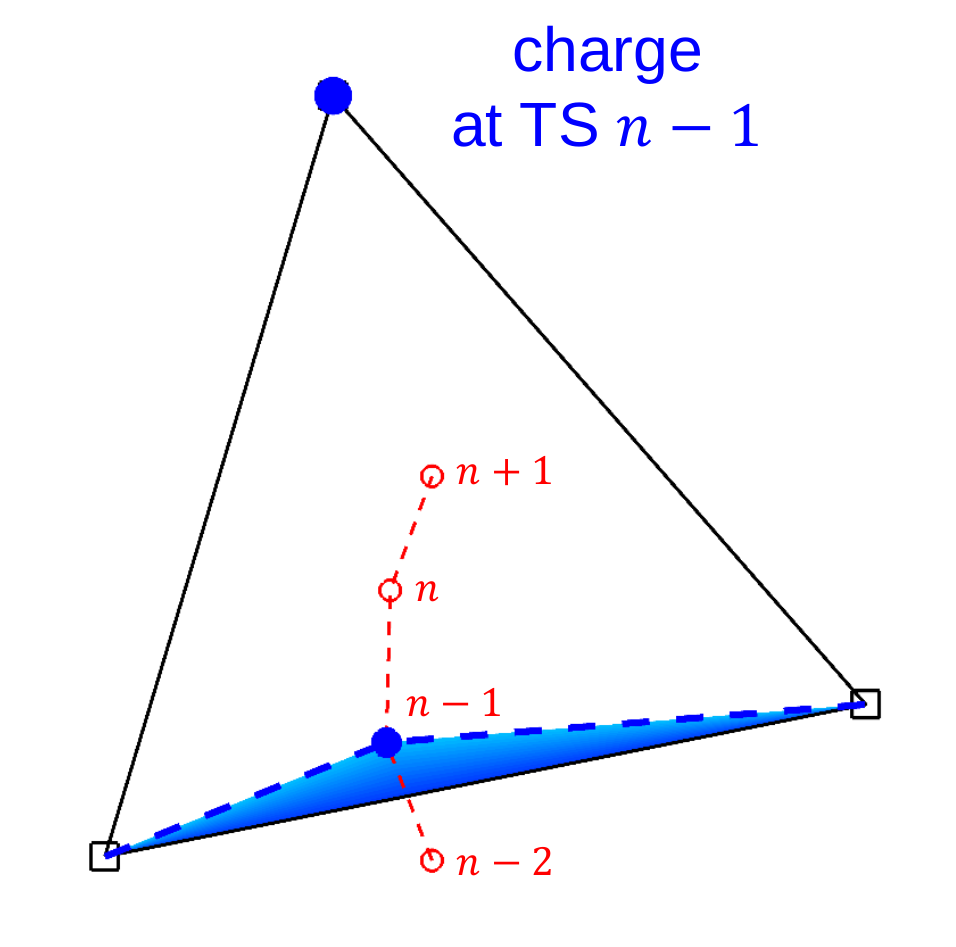}
    }
	\subfloat[\label{fig:q_def_c}]{
    \includegraphics[width=1.1in]{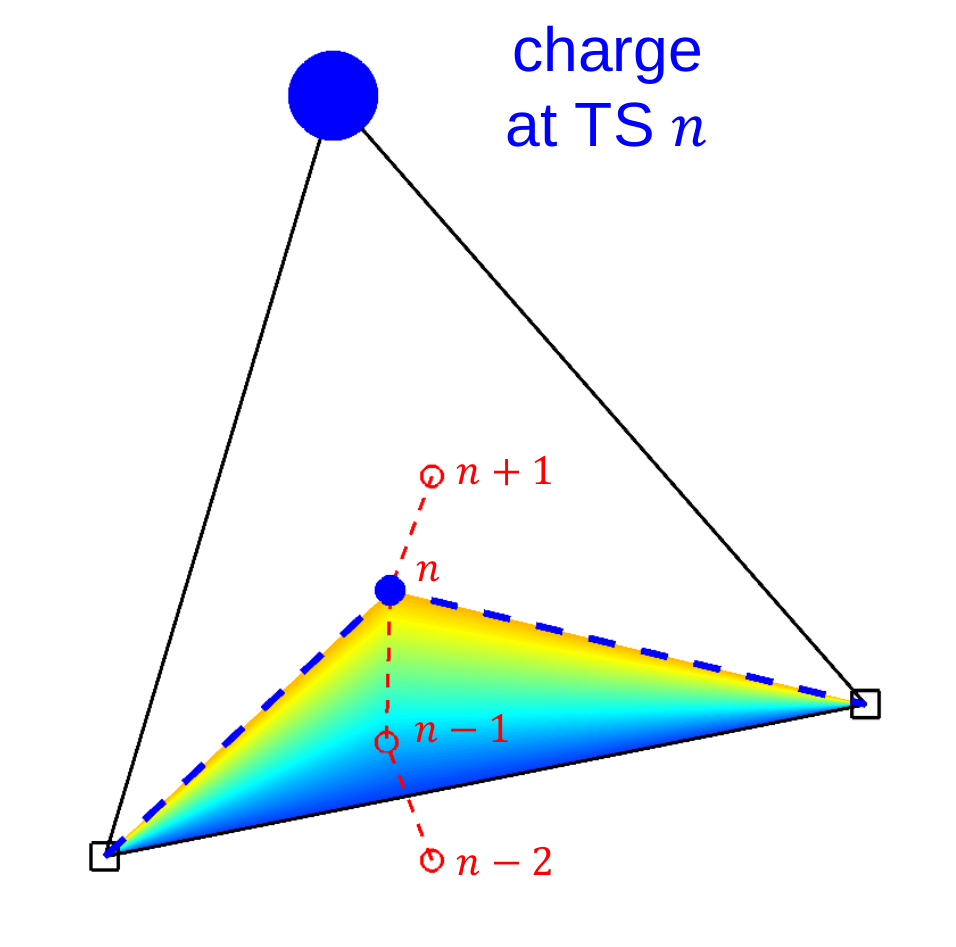}
    }
	\subfloat[\label{fig:q_def_d}]{
    \includegraphics[width=1.1in]{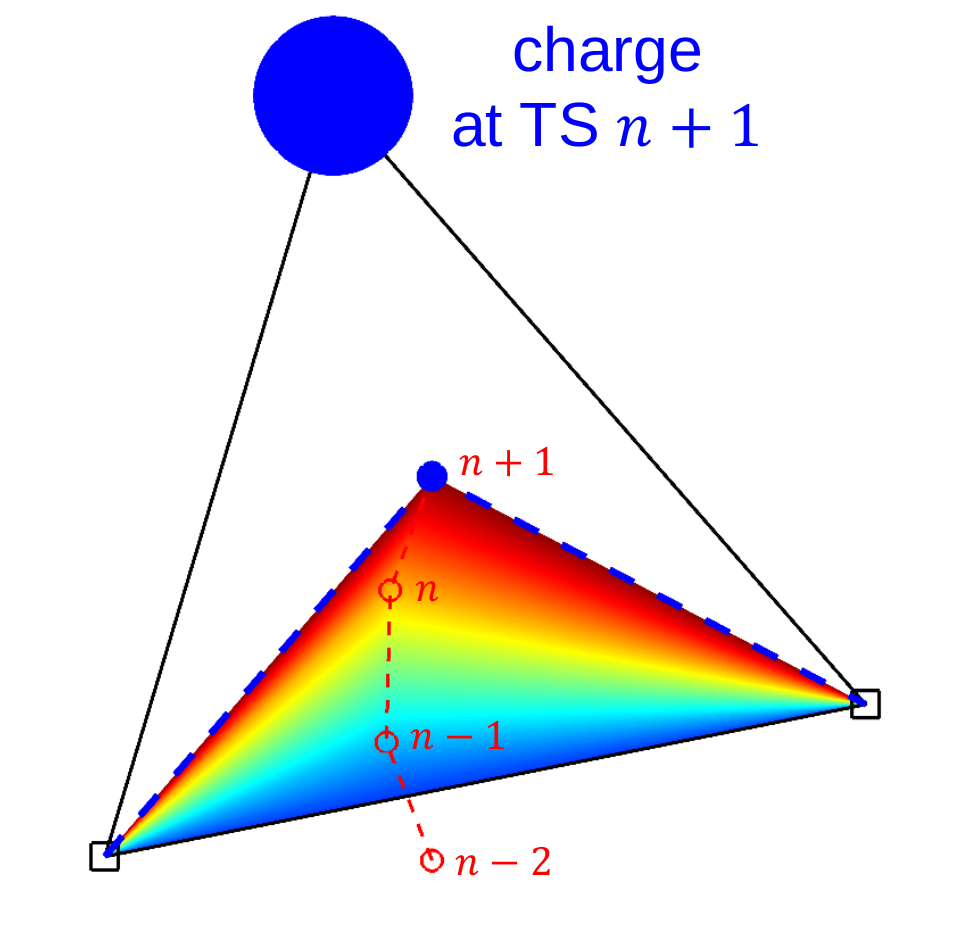}
    }
\caption{The grid charge evolution at node $1$ during a single particle push. The charge states at time-steps of $n-2$, $n-1$, $n$, and $n+1$ are illustrated in (a), (b), (c), and (d), respectively. Note that a colored triangular area, formed by the particle position, nodes $2$, $3$, corresponds to the grid charge of node $1$ marked as a blue-colored circle.}
\label{fig:q_def}
\end{figure}
}

\subsection{Charge-conserving particle scatter} \label{sec:charge-conserving particle scatter}
The particle scatter transfers the kinetic information from the charged particles to the grid for its subsequent use by the field solver.
As described before, the discrete (Hodge dual) charge and current densities are defined at nodal and edge elements of the primal mesh, respectively.
As mentioned in Sec. \ref{sec:introduction}, conventional interpolation techniques for irregular grids may violate charge conservation and cause spurious (nonphysical) charge deposition on the grid~\cite{pinto2014charge,na2016local}.
A charge-conserving scatter algorithm has been recently proposed~\cite{moon2015exact} based on the use of discrete Whitney forms as a consistent interpolation (``particle shape") factor. 
For piecewise-linear particle trajectories, charges and currents associated with 0- and 1-forms are defined as follows~\cite{moon2015exact}:
\begin{eqnarray}
\!\!\!\!\!\!\!\!\!\!\!\!\!\!
\mathbb{Q}_{\star,i}^{n}
\!\!\!\!&\equiv&\!\!\!\!
\sum_{\alpha}q_{\alpha}
W_{i}^{0}\left(\vec{r}_{\alpha}^{n}\right),
=\sum_{\alpha}q_{\alpha}
\lambda_{i,\alpha}^{n},
\label{eq:charge}
\\
\!\!\!\!\!\!\!\!\!\!\!\!\!\!
\mathbb{J}_{\star,i}^{n+\frac{1}{2}}
\!\!\!\!&\equiv&\!\!\!\!
\sum_{\alpha}\frac{q_{\alpha}}{\Delta_{t}}
\int_{\vec{r}^{n}_{\alpha}}^{\vec{r}_{\alpha}^{n+1}}\vec{W}^{1}_{i}\left(\vec{r}\right)\cdot d\vec{r}
=
\sum_{\alpha}\frac{q_{\alpha}}{\Delta_{t}}
\left(
\lambda_{i_{a},\alpha}^{n+1}\lambda_{i_{b},\alpha}^{n}
-
\lambda_{i_{b},\alpha}^{n+1}\lambda_{i_{a},\alpha}^{n}
\right),
\label{eq:current}
\end{eqnarray}
where
$i$ represents the $i$-th grid element, and $i_{a}$ and $i_{b}$ denote indices of lower dimensional manifolds of $i$-th element\footnote{For example, if $i$-th element is an edge, $i_{a}$ and $i_{b}$ indicate the two nodes associated with the said edge.}. The functions $W_{i}^{0}$ and $\vec{W}_{i}^{1}$ are scalar and vector proxies of Whitney 0- and 1-forms, respectively, and $\lambda_{i,\alpha}^{n}$ is a barycentric coordinate for an $\alpha$-species particle at time-step $n$, respectively.
One can find the derivations of Eqs. (\ref{eq:charge}) and (\ref{eq:current}) in more detail in~\cite{moon2015exact}.
\begin{figure}[t]
    \centering
	\subfloat[\label{fig:j_def_a}]{
    \includegraphics[width=1.1in]{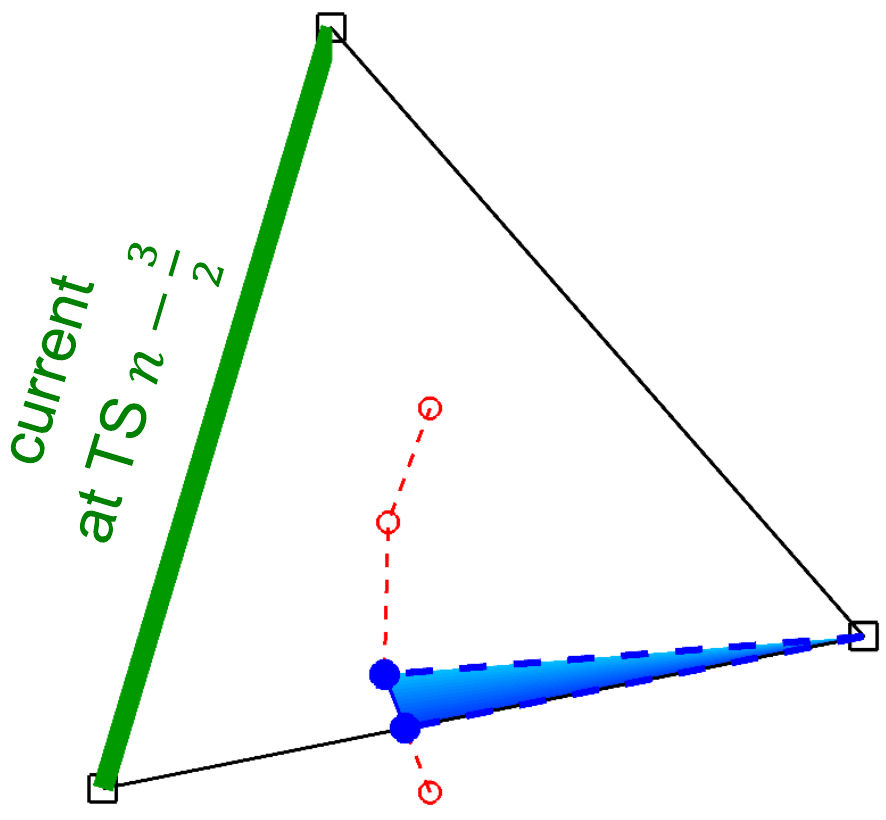}
    }
	\subfloat[\label{fig:j_def_b}]{
    \includegraphics[width=1.1in]{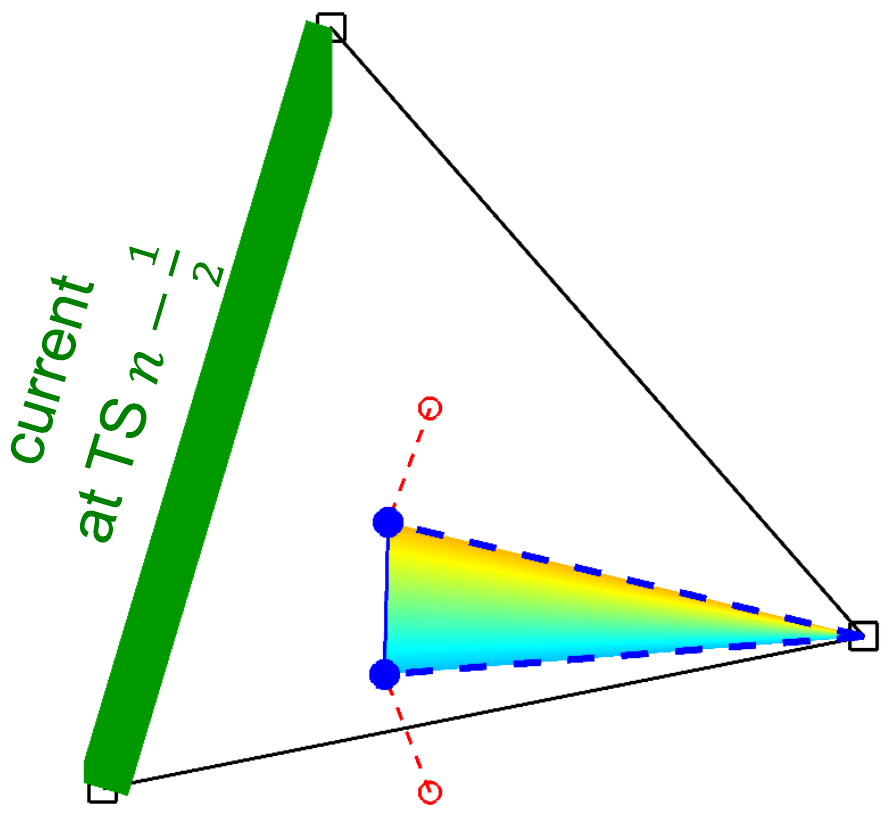}
    }
	\subfloat[\label{fig:j_def_c}]{
    \includegraphics[width=1.1in]{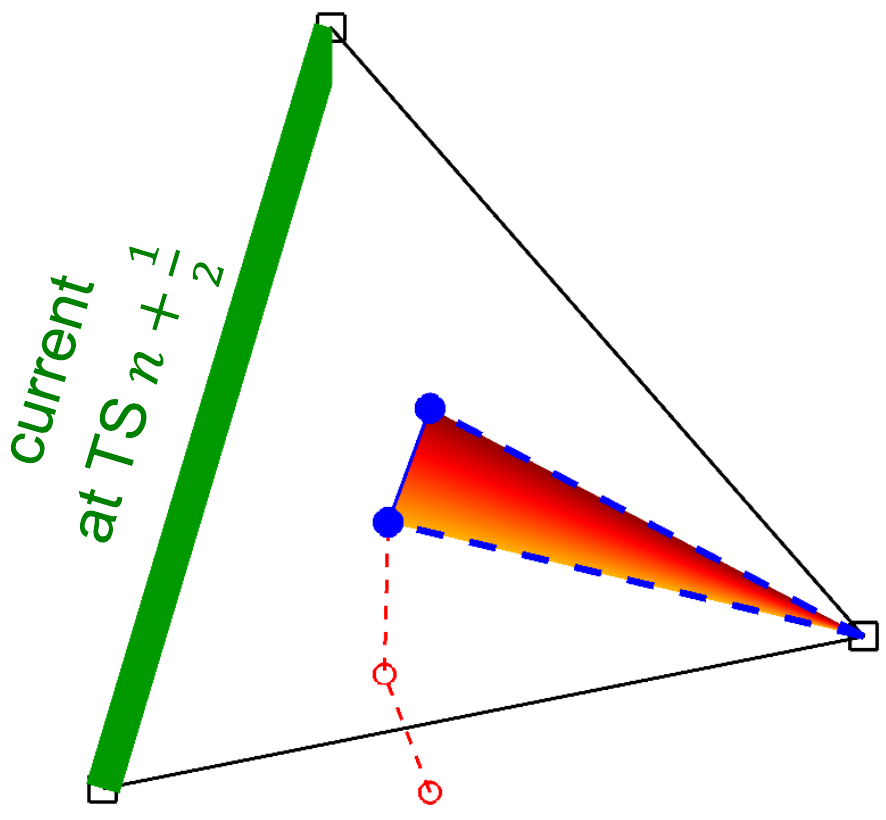}
    }
    \caption{The grid current evolution at edge $1$ is indicated by the thickness of the green line. Similar to the nodal charge, its equivalent amount is illustrated by a colored triangular area.}
\label{fig:j_def}
\end{figure}
Fig. \ref{fig:q_def} and Fig. \ref{fig:j_def} provide the geometrical depiction on the node-based charges and edge-based currents, respectively,
obtained by Eqs. (\ref{eq:charge}) and (\ref{eq:current}).
For example, assume a charged particle movement from time-step, $n-2$ to time-step, $n+1$. 
As shown in Fig. \ref{fig:q_def}, the grid charge (blue-colored circle) at node $1$ is calculated based on (\ref{eq:charge}). 
The associated charge value is represented by the colored triangular area formed by the particle position and nodes $2$ and $3$.
Likewise, as shown in Fig. \ref{fig:j_def}, the respective grid current at edge $1$ (\ref{eq:current}) is given by a gradient-colored triangular area formed by the particle trajectory and the node that does not belong to that edge.

\subsection{Verification of charge conservation}\label{sec:Verification of charge conservation}
\subsubsection{Discrete continuity equation (DCE)}
In order to verify  charge conservation, we examine the discrete charge continuity equation (DCE) given by~\cite{moon2015exact,na2016local}
\begin{eqnarray}
\frac{
\left[\mathbb{Q}\right]^{n+1}-\left[\mathbb{Q}\right]^{n}
}{\Delta t}
+
\left[\tilde{\mathcal{D}}_{\text{div}}\right]
\cdot
\left[\mathbb{J}\right]^{n+\frac{1}{2}}
=
0
\label{eq:DCE}
\end{eqnarray}
where $\left[\tilde{\mathcal{D}}_{\text{div}}\right]$ is an incidence matrix encoding the discrete representation of the exterior derivative in the dual mesh or, equivalently, the divergence operator distilled from metric information. Incidence matrices take their elements in the set of $\left\{-1,0,1\right\}$, see also the Appendices.
Note that Hodge matrices for the discrete Hodge duals of charges and currents ($\left[\mathbb{Q}_{\star}\right]$ and $\left[\mathbb{J}_{\star}\right]$) become identity matrices for barycentric dual lattices. Therefore, $\left[\mathbb{Q}\right]=\left[\mathbb{Q}_{\star}\right]$ and $\left[\mathbb{J}\right]=\left[\mathbb{J}_{\star}\right]$.
\begin{figure}[t]
\centering
\includegraphics[width=3.3in]{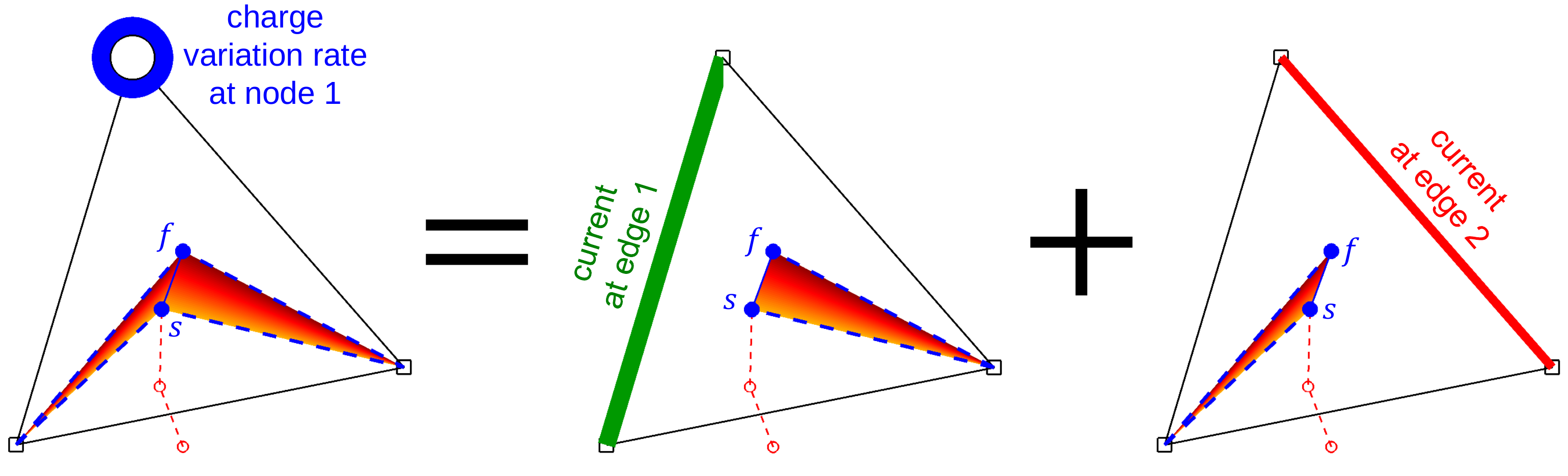}
\caption{Schematics for verifying charge conservation. The charge rate at node $1$ exactly matches the net current flowing from/to node $1$.}
\label{fig:DCE_schematic}
\end{figure}
The first term in the left-hand side of (\ref{eq:DCE}) stands for the rate of change in the charge during one time-step. The second term in the left-hand side of (\ref{eq:DCE}) represents the net current flowing from/to the nodes.
Consider a charged particle that moves from time-step of $n$ to time-step $n+1$ in a given triangular element (see Fig. \ref{fig:DCE_schematic}).
Then, the rate of change of the nodal charge at node $1$ is 
\begin{eqnarray}
\frac{\mathbb{Q}_{\star,1,\alpha}^{n+1}-\mathbb{Q}_{\star,1,\alpha}^{n}}
{\Delta{t}}
=
\frac{q_{\alpha}}{\Delta{t}}
\left(
\lambda_{1,\alpha}^{n+1}-\lambda_{1,\alpha}^{n}
\right).
\end{eqnarray}
The net current flowing from/to node $1$ is a sum of the edge currents at edges $1$ and $2$ as follows:
\begin{eqnarray}
\mathbb{J}_{\star,1,\alpha}^{n+\frac{1}{2}}+\mathbb{J}_{\star,2,\alpha}^{n+\frac{1}{2}}
\!\!\!\!&=&\!\!\!\!
\frac{q_{\alpha}}{\Delta{t}}
\left(
\lambda_{1,\alpha}^{n+1}\lambda_{2,\alpha}^{n}
-
\lambda_{2,\alpha}^{n+1}\lambda_{1,\alpha}^{n}
\right)
+
\frac{q_{\alpha}}{\Delta_{t}}
\left(
\lambda_{1,\alpha}^{n+1}\lambda_{3,\alpha}^{n}
-
\lambda_{3,\alpha}^{n+1}\lambda_{1,\alpha}^{n}
\right)
\nonumber\\
\!\!\!\!&=&\!\!\!\!
-\frac{q_{\alpha}}{\Delta_{t}}
\left(
\lambda_{1,\alpha}^{n+1}-\lambda_{1,\alpha}^{n}
\right).
\end{eqnarray}
Thus, the rate of change of the nodal charge at node $1$ is equal to the net current flow.
From the geometrical perspective, this is depicted in Fig. \ref{fig:DCE_schematic}.

\subsubsection{Discrete Gauss' law (DGL)}
Premultiplying both sides of (33) by $\left[\tilde{\mathcal{D}}_{\text{div}}\right]\cdot\left[\star_{\epsilon}\right]_{\text{a}}$, where
 $\left[\tilde{\mathcal{D}}_{\text{div}}\right]$ is the incidence matrix representing the discrete divergence operator on the dual grid and $\left[\star_{\epsilon}\right]_{\text{a}}$ is an exact inverse of $\left[\star_{\epsilon}\right]]_{\text{a}}^{-1}$ obtained by using SPAI algorithm, and utilizing the identity $\left[\tilde{\mathcal{D}}_{\text{div}}\right]\cdot\left[\tilde{\mathcal{D}}_{\text{curl}}\right]=0$~\cite{teixeira1999lattice,teixeira2014lattice} yields
\begin{eqnarray}
\left[\tilde{\mathcal{D}}_{\text{div}}\right]\cdot\left[\star_{\epsilon}\right]_{a}\cdot\left[\mathbb{E}\right]^{n+1}
=
\left[\tilde{\mathcal{D}}_{\text{div}}\right]\cdot\left[\star_{\epsilon}\right]_{a}\cdot\left[\mathbb{E}\right]^{n}
-{\Delta t}\left[\tilde{\mathcal{D}}_{\text{div}}\right]\cdot\left[\mathbb{J}\right]^{n+\frac{1}{2}}.
\label{eq:md_A_2}
\end{eqnarray}
which can be rearranged as 
\begin{eqnarray}
\left[\tilde{\mathcal{D}}_{\text{div}}\right]\cdot\left[\star_{\epsilon}\right]_{a}\cdot\left(\frac{\left[\mathbb{E}\right]^{n+1}-\left[\mathbb{E}\right]^{n}}{\Delta t}\right)
=
-\left[\tilde{\mathcal{D}}_{\text{div}}\right]\cdot\left[\mathbb{J}\right]^{n+\frac{1}{2}},
\end{eqnarray}
and, using (\ref{eq:DCE}), rewritten as
\begin{eqnarray}
\left[\tilde{\mathcal{D}}_{\text{div}}\right]\cdot\left[\star_{\epsilon}\right]_{a}\cdot\left(\frac{\left[\mathbb{E}\right]^{n+1}-\left[\mathbb{E}\right]^{n}}{\Delta t}\right)
=
\frac{\left[\mathbb{Q}\right]^{n+1}-\left[\mathbb{Q}\right]^{n}}{\Delta t}.
\label{eq:DGL}
\end{eqnarray}
Eq. (\ref{eq:DGL}) implies that residuals of the discrete Gauss' law (DGL) at any two successive time steps remain the same, in other words
\begin{eqnarray}
\left[\tilde{\mathcal{D}}_{\text{div}}\right]\cdot\left[\star_{\epsilon}\right]_{a}\cdot\left[\mathbb{E}\right]^{n+1}-\left[\mathbb{Q}\right]^{n+1}
=
\left[\tilde{\mathcal{D}}_{\text{div}}\right]\cdot\left[\star_{\epsilon}\right]_{a}\cdot\left[\mathbb{E}\right]^{n}-\left[\mathbb{Q}\right]^{n},\
\end{eqnarray}
and by induction,
\begin{eqnarray}
\underbrace{
\left[\tilde{\mathcal{D}}_{\text{div}}\right]\cdot\left[\star_{\epsilon}\right]_{a}\cdot\left[\mathbb{E}\right]^{n}-\left[\mathbb{Q}\right]^{n}
}_{\text{res}^{n}}
=
\underbrace{
\left[\tilde{\mathcal{D}}_{\text{div}}\right]\cdot\left[\star_{\epsilon}\right]_{a}\cdot\left[\mathbb{E}\right]^{0}-\left[\mathbb{Q}\right]^{0}
}_{\text{res}^{0}}
\label{DGL_RSD_REL}
\end{eqnarray}
for all $n$. This means that if consistent initial conditions $\left[\tilde{\mathcal{D}}_{\text{div}}\right]\cdot\left[\star_{\epsilon}\right]_{a}\cdot\left[\mathbb{E}\right]^{0}=\left[\mathbb{Q}\right]^{0}$ are set, then the DGL remains valid for all time steps.

\section{Validation}\label{sec:Validation}
In this section, we provide validation examples. First, we consider a metallic cylindrical cavity and compare the resonant frequencies of the $\text{TM}_{0np}$ cavity modes obtained by the present field solver with the exact (analytic) results. Second, we model a space-charge-limited cylindrical diode with a finite-length emitter and compare the maximum injection currents for divergent and convergent electron beam flows against previously published results.

\subsection{Metallic hollow cylindrical cavity}
Assume a hollow cylindrical cavity with metallic walls, radius $a = 0.5$ m and height $h = 1$ m (see Fig. \ref{fig:cir_cavity_field_geom}). 
Two magnetic point sources $M_{1}\left(\rho,z,t\right)$ and $M_{2}\left(\rho,z,t\right)$ excited by Gaussian-modulated pulses of broadband spectrum are placed at arbitrary
positions given as
\begin{flalign}
M_{1}\left(\rho=0.37,z=0.2,t\right)=&0.7e^{-\left(\frac{t-t_{g}/2}{2\sigma_{g}}\right)^{2}}\cos{\left[\omega_{g}\left(t-t_{g}/2\right)\right]}
\\
M_{2}\left(\rho=0.08,z=0.7,t\right)=&0.5 e^{-\left(\frac{t-t_{g}}{2\sigma_{g}}\right)^{2}}\cos{\left[\omega_{g}\left(t-t_{g}\right)\right]}
\end{flalign}
with $t_{g}=40$ ns, $\sigma_{g}=1$ ns, and  $\omega_{g}=\pi \times 10^{9}$ rad/s.
The source locations are denoted by Tx in Fig. \ref{fig:cir_cavity_field_geom}. These sources excite resonant modes inside the cavity. 
The 2D meridian plane of the cylindrical cavity is discretized by an unstructured grid with $8,095$ nodes, $23,928$ edges , and $15,834$ faces.
The three lateral metallic boundaries are assumed as perfect electric conductors (PEC). The remaining boundary is the $z$-axis (axisymmetric boundary).
The time step interval $\Delta {t}$ is chosen as $1$ ps, and the simulation runs over a total of $2\times10^{6}$ time steps.
Fig.~\ref{fig:cir_cavity_field_geom} shows a snapshot for electric field distribution in the cavity at $2$ $\mu$s. 
The RGB colormap and the white arrows indicate magnitudes and vectors of the electric fields, respectively.
\begin{figure}[t]
	\centering
	\includegraphics[width=3in]{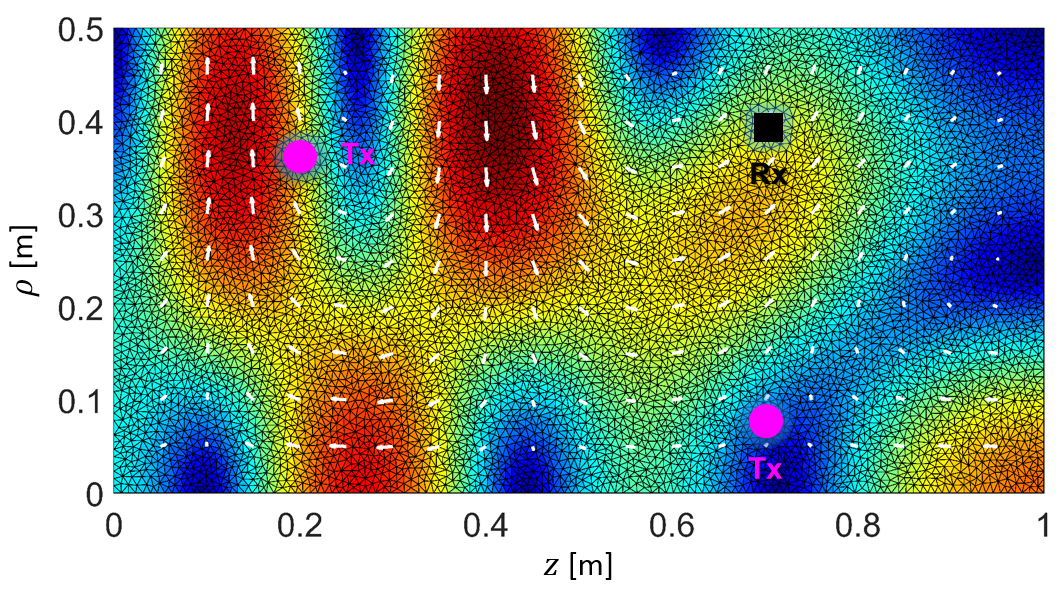}
	\caption{Snapshots for electric field distribution at 2 $\mu$s. Note that RGB colors and white arrows indicate magnitudes and vectors of the electric fields, respectively.}
	\label{fig:cir_cavity_field_geom}
\end{figure}
The time signal are detected at $\left(\rho=0.71,z=0.39\right)$ (Rx in Fig. \ref{fig:cir_cavity_field_geom}) and a Fourier analysis is performed to obtain a spectrum of the signal.
The resulting spectrum shows the resonant cavity modes from $1$ MHz to $1$ GHz in Fig. \ref{fig:cavity_spectrum}, where blue-solid lines are axisymmetric FETD field-solver results and red-dashed lines are analytic results. An excellent agreement can be observed. 
In addition, Table.~\ref{tab:spectrum_cavity} shows the resonant frequencies for $\text{TM}_{mnp}$ cavity modes\footnote{$m$, $n$, and $p$ are associated with eigenmode orders along azimuthal ($\phi$), radial ($\rho$), and longitudinal directions ($z$), respectively.} and the normalized error defined as $\frac{\left|f_{s}-f_{a}\right|}{f_{a}}$ where $f_{s}$ and $f_{a}$ are numerical and analytic resonant frequencies, respectively.  It is seen that all resonant cavity modes are axisymmetric ($m=0$) and the normalized errors are fairly low.

\begin{figure}[t]
	\centering
	\includegraphics[width=3in]{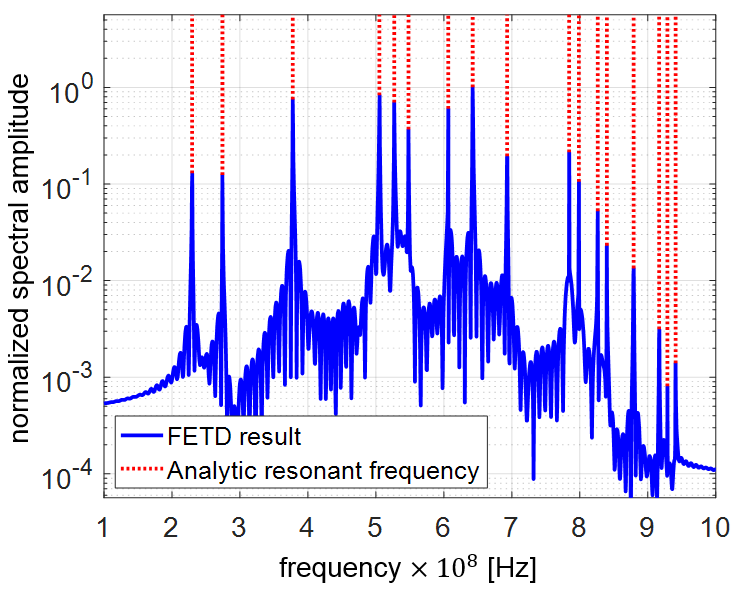}
	\caption{Spectrum for resonant cavity modes from $1$ MHz to $1$ GHz.}
	\label{fig:cavity_spectrum}
\end{figure}

\begin{table} [t]
\caption{Resonant frequencies for axisymmetric cavity modes and normalized errors between numerical and analytic works.}  
\centering
{\scriptsize %
\begin{tabular}{c c c c c c}
\hline\hline 
\\[-0.5em]
$\left(m,n,p\right) $ & $f_{a}$ [GHz] & $\frac{\left|f_{a}-f_{s}\right|}{f_{a}}\times 100$ [\%] & $\left(m,n,p\right) $ & $f_{a}$ [GHz] & $\frac{\left|f_{a}-f_{s}\right|}{f_{a}}\times 100$ [\%] \\
\\[-0.5em]
\hline 
\\[-0.5em]
$\left(0,1,0\right)$ & $0.230$ & $0.083$ & $\left(0,1,5\right)$ & $0.784$ & $0.0035$ \\
$\left(0,1,1\right)$ & $0.274$ & $0.040$ & $\left(0,2,4\right)$ & $0.799$ & $0.0048$ \\
$\left(0,1,2\right)$ & $0.378$ & $0.040$ & $\left(0,3,0\right)$ & $0.826$ & $0.0014$ \\
$\left(0,1,3\right)$ & $0.505$ & $0.047$ & $\left(0,3,1\right)$ & $0.840$ & $0.019$ \\
$\left(0,2,0\right)$ & $0.527$ & $0.042$ & $\left(0,3,1\right)$ & $0.879$ & $0.017$ \\
$\left(0,2,1\right)$ & $0.548$ & $0.031$ & $\left(0,3,1\right)$ & $0.917$ & $0.027$ \\
$\left(0,2,2\right)$ & $0.607$ & $0.0030$ & $\left(0,2,5\right)$ & $0.929$ & $0.027$ \\
$\left(0,1,4\right)$ & $0.642$ & $0.023$ & $\left(0,3,3\right)$ & 0.941 & 0.016 \\
$\left(0,2,3\right)$ & $0.693$ & $0.034$ & $-$ & $-$ & $-$ \\
\\[-0.75em]
\hline
\end{tabular}
}
\label{tab:spectrum_cavity}
\end{table}

\subsection{Space-charge-limited (SCL) cylindrical diode}
As a second validation example, we test the accuracy of present EM-PIC algorithm by modeling a SCL cylindrical diode with finite-length emitter.
By applying an external voltage to the cathode, a fast rise in the number of electrons emitted from the cathode initially occurs; however, in the steady-state the injection current density eventually becomes saturated due to space charge effects.
For an infinitely long cylindrical diode and electrodes, Langmuir-Blodgett's law describes the SCL current per unit length $J_{1D,LB}$ as
\begin{flalign}
J_{1D,LB}\equiv\frac{8\pi\epsilon_{0}}{9}\sqrt{\frac{2e}{m}}\frac{V^{3/2}}{\rho\beta}
\end{flalign}
where $V$ is the external voltage, $e$ and $m$ are charge and mass of electrons, $\rho$ is a radial coordinate, and 
\begin{flalign}
\beta=\mu-\frac{2\mu^{2}}{5}+\frac{11\mu^{3}}{120}-\frac{47\mu^{4}}{3300}+...,
\end{flalign}
with $\mu=\ln\left(\rho/\rho_{c}\right)$ and $\rho_{c}$ denotes the radius of the cathode.
Here, we consider the emitting electrode with finite length $L_{e}$.
The solution domain is $\Omega=\left\{\left(\rho,z\right)\in\left[\rho_{i},\rho_{o}\right]\times\left[0,L_{z}\right]\right\}$ with $\rho_{i}=5$ mm, $\rho_{o}=15$ mm, and $L_{z}=100$ mm. The domain has the horizontal wall segments representing electrode surfaces (cathode or anode), as shown in Fig. \ref{fig:cd_dcv_flow}.
If the inner conductor is chosen as a cathode, the electron flow becomes divergent, on the other hand, it is convergent. 
The left and right boundaries of the domain are truncated by a perfectly matched layer (PML)~\cite{teixeira1998general,teixeira1999unified,WangTPS2006}.
The unstructured mesh has $7,313$ nodes, $21,334$ edges, and $14,022$ faces. We choose $\Delta t=0.15$ ps and the simulation runs up to a total $30,000$ time-steps.

\begin{figure}[t]
\centering
\subfloat[\label{fig:cd_dv_flow}]{
	\includegraphics[width=4.5in]{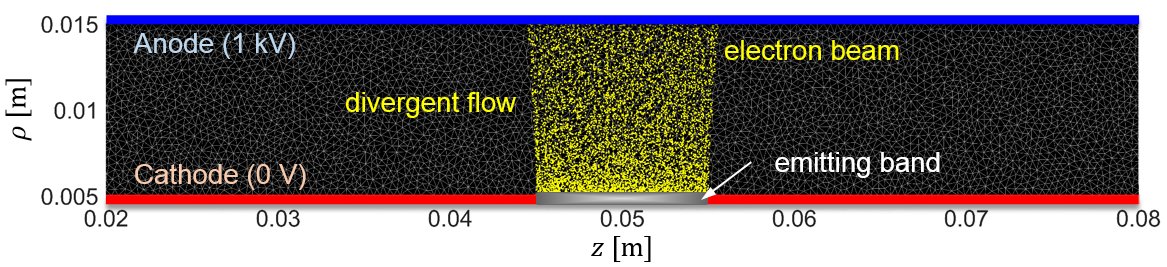}
}
	\\
\subfloat[\label{fig:cd_cv_flow}]{
	\includegraphics[width=4.5in]{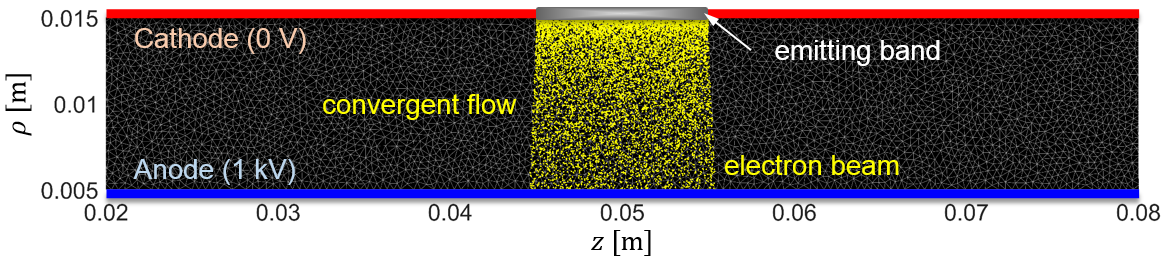}
}
\caption{Schematics for divergent and convergent flows in the cylindrical diode.}
\label{fig:cd_dcv_flow}
\end{figure}
Fig. \ref{fig:cd_dcv_flow} illustrates snapshots for electron beam distribution with $L_{e}=10$ mm and $V=1$ kV.
In order to determine $J_{2D,\text{max injection}}$, we first fix the superparticle scaling factor $C_{sp}$, indicating the number of electrons for each superparticle, and gradually increase the superparticle injection rate until the virtual cathode starts to form.
We simulate a total of $8$ cases, including $L_{e}/\rho_{o}=0.4$, $1$, $2.425$, and $4$ for each divergent and convergent electron flow, and compare the maximum injection current density without formation of a virtual cathode, $J_{2D,\text{max injection}}$ to the previous results obtained by~\cite{kostov2002space} with \texttt{KARAT}, which is a FDTD-based EM-PIC algorithm. 
Fig. \ref{fig:scl_cd_comp} shows $J_{2D,\text{max injection}}$ versus $L_{z}/\rho_{o}$ for divergent and convergent electron flows.
Red-solid (divergent) and -dashed (convergent) lines are \texttt{KARAT} results and blue-markers with upper ranges are obtained from present EM-PIC simulations.
The upper ranges on the blue markers stand for an interval where an exact solutions for the current density may exist.
The marker indicates maximum current density without the virtual cathode formation and upper horizontal line is minimum injection current density with virtual cathode formation.
Two markers are used because of the stepwise increases in the current density by the assumed superparticle number in our EM-PIC model.
Decreasing superparticle scaling factor or increasing superparticle injection rate can yield higher resolution.
Fig. \ref{fig:scl_cd_comp} shows very good agreement between the results of the present EM-PIC algorithm and those of \texttt{KARAT}, for both divergent and convergent flows.
In addition, it is seen that, as $L_{e}/\rho_{o}$ increases, the current density converges to the limit of Langmuir-Blodgett's law as expected.
\begin{figure}[t]
	\centering
	\includegraphics[width=3.2in]{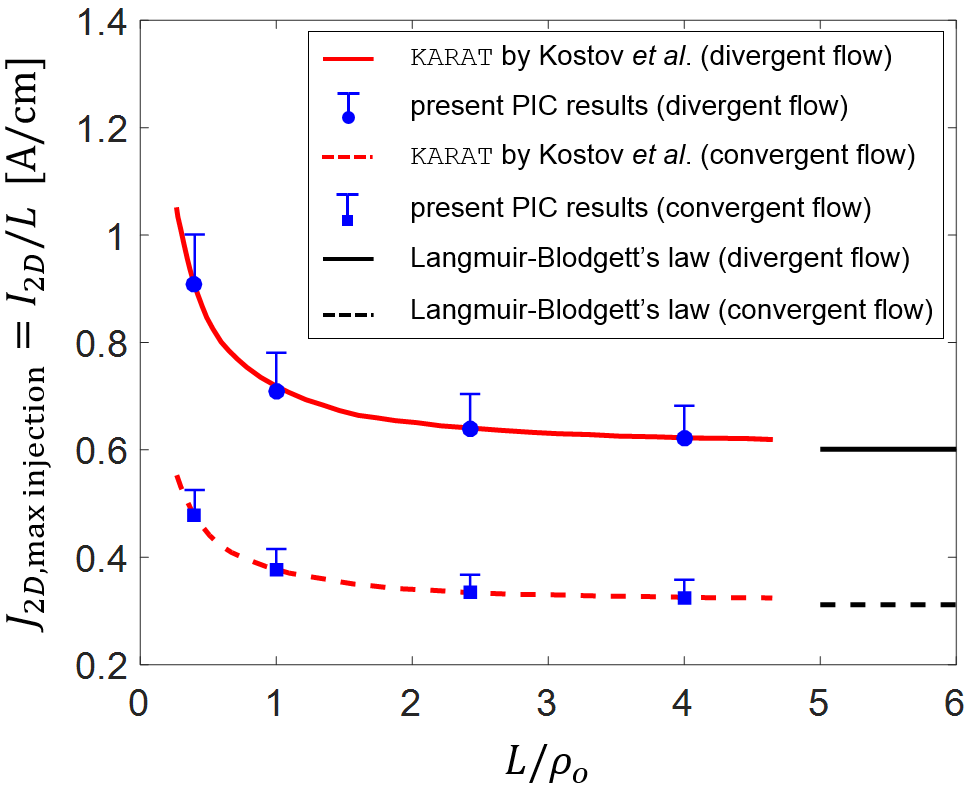}
	\caption{Space-charge-limited current density for various $L_{z}/\rho_{o}$ and comparison between present EM-PIC simulations and \texttt{KARAT} by \cite{kostov2002space}.}
	\label{fig:scl_cd_comp}
\end{figure}

Fig. \ref{fig:vcf} shows the magnitude of the electrical self-field and external field at the instant of virtual cathode formation.
Since the external field is stronger as $\rho$ decreases, a divergent flow produces more charges on the cathode surface so that their self-field may cancel the external field. As a result, the current density of divergent flows is larger than that of convergent flow.
\begin{figure}[t]
\centering
\subfloat[\label{fig:dv_vcf}]{
	\includegraphics[width=4.5in]{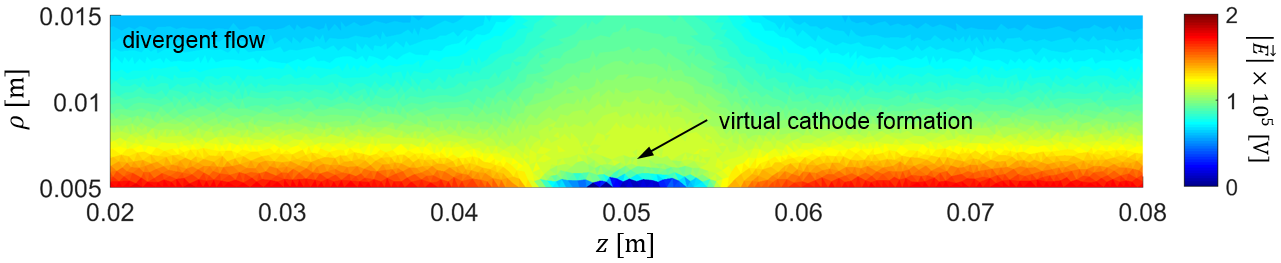}
}
	\\
\subfloat[\label{fig:cv_vcf}]{
	\includegraphics[width=4.5in]{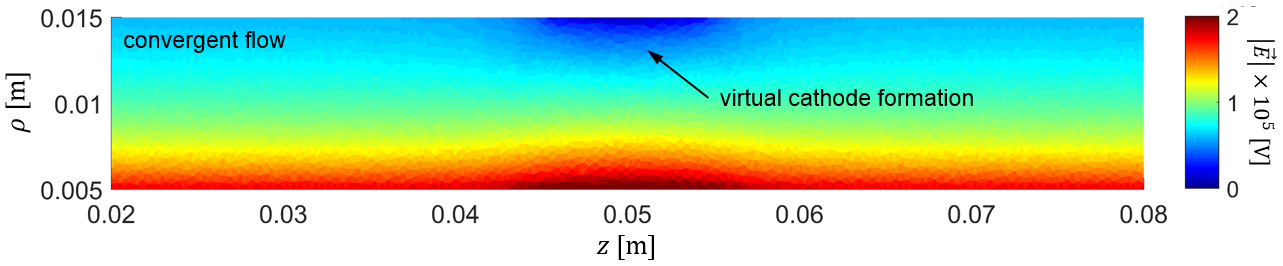}
}
\caption{Electric field intensity of self- and external fields at the instant of virtual cathode formation.}
\label{fig:vcf}
\end{figure}

\section{Numerical examples}\label{sec:Numerical examples}
In this section, we present simulations of a relativistic BWO device operating at $\pi$-point by using the proposed EM-PIC algorithm.
First we consider a SWS with sinusoidal ripples on a cylindrical waveguide section and determine its characteristics by performing a ``cold" test (i.e. without the presence of an electron beam). 
Then, we perform ``hot" tests (with electron beams) of BWO to check the reliability and validity of our axisymmetric EM-PIC algorithm. 
In particular, we are interested in investigating effects of staircasing errors on the predicted behavior of the BWO system.

\begin{figure}[t]
	\centering
	\includegraphics[width=4.5in]{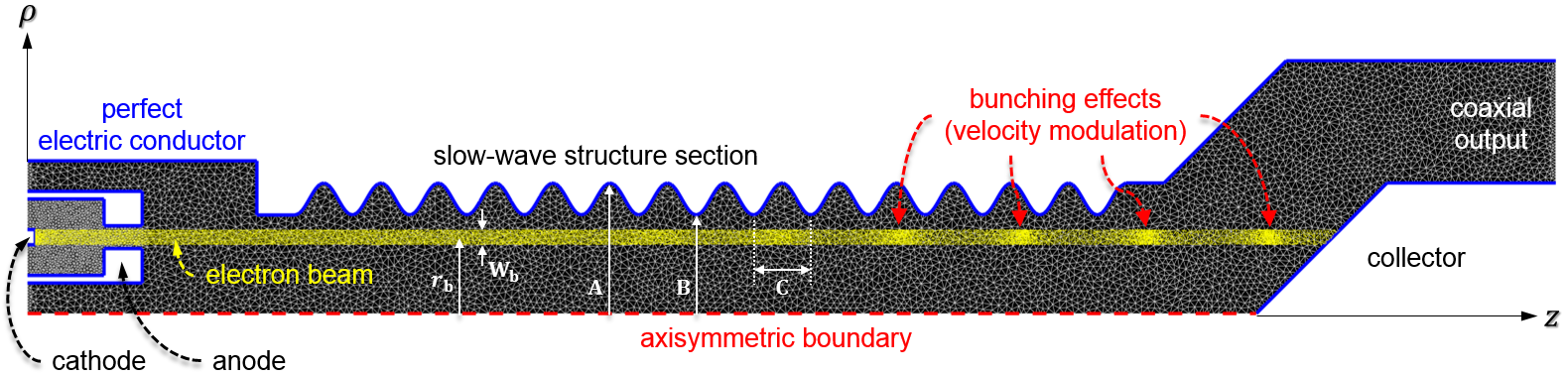}
	\caption{Schematics of backward-wave oscillator with an instant particle distribution snapshots at $t=21.50$~ns.}
	\label{fig:particle_distribution}
\end{figure}
\begin{figure}[t]
	\centering
	\includegraphics[width=2.5in]{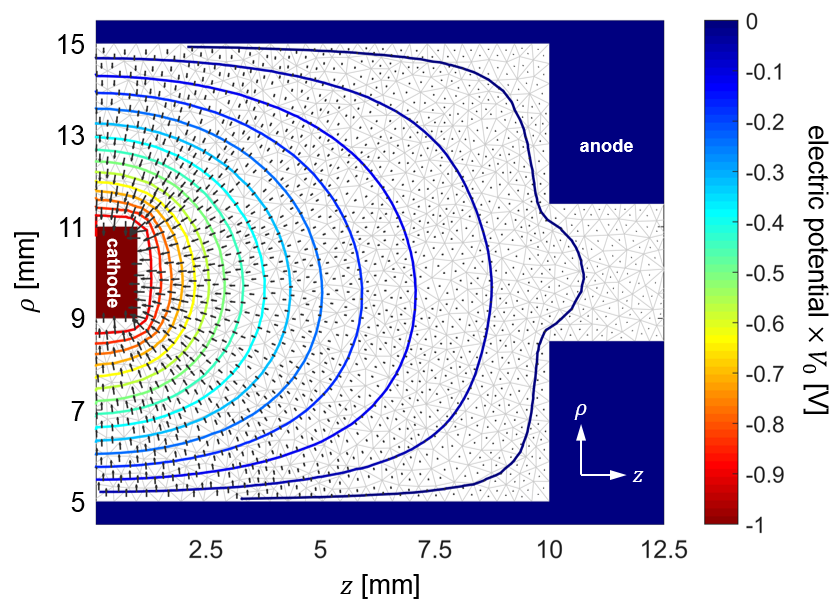}
	\caption{Electric potential distribution (contour plots) and corresponding electric fields (vector plots) between the cathode and the anode.}
	\label{fig:poisson_solver}
\end{figure}

\subsection{Relativistic backward-wave oscillator (BWO)}\label{sec:Relativistic backward-wave oscillator}
Consider a BWO system composed of the cathode-anode, the SCSWS region, the beam collector, the output port, and the beam focusing system.
In order to produce the relativistic electron beam, we apply an external voltage difference  $V_{0}$ between cathode and anode.
We choose $-550$ kV which produces the electron beam with mean axial velocity $v_{\text{beam}}=0.877c$ with the width of $2$ mm.
Fig. \ref{fig:poisson_solver} illustrates electric potential distribution (contour plots) and corresponding electric fields (vector plots) by solving Poisson equations.
Each super-particle represents $1.495 \times10^{7}$ electrons so that the total injection current is around $1.5$ kA. On average, $102$ macro-particles are assigned to each grid cell.
The Debye length $\lambda_{D}$ in our simulations is $10.7258$~mm and much larger than the average grid (edge) length, $l_{av}=1.1123$~mm. This avoids artificial heating of the electron beam.
The electron beam is emitted from the cathode and eventually absorbed at the collector. 
We consider a SCSWS with radial profile $R(z)=\frac{1}{2}(A+B) + \frac{1}{2}(A-B)\cos\left(\frac{2\pi}{C} z\right)$, where $A$ and $B$ are maximum and minimum  radii, respectively, and $C$ is the corrugation period. 
The total number of corrugations along the structure is denoted as $N_{\text{crg}}$.
Based on an eigenmode analysis~\footnote{Corresponding to the absence of an electron beam or a so-called `cold test'.} (see the double-refined case in Fig. \ref{fig:dispersion_comp}), the SCSWS was designed to have $A=17.1$~mm, $B=12.9$~mm, $C=7.5$~mm, and $N_{\text{crg}}=14.5$ for Ku band operation. 
We terminate the output ports of the BWO system by inserting a PML. All left vertical walls except for the cathode are truncated by PML to avoid spurious reflections.
By using a PML with thickness equals to $0.2\lambda_{0}$ where $\lambda_{0}$ is wavelength of the center frequency, the PML reflection coefficient, defined in \cite{donderici2008mixed}, is as low as $-92$ dB.
In the beam focusing system, a static axial magnetic field is applied over the SWS region.
We enforce axisymmetric fields at the $z$-axis, by applying a perfect magnetic conductor (PMC) boundary condition there (note that only the $E_{z}$ component is present on axis). For the particles incident on the axis we use a perfectly reflecting boundary condition. 
\begin{figure}[t]
	\centering
	\includegraphics[width=4in]{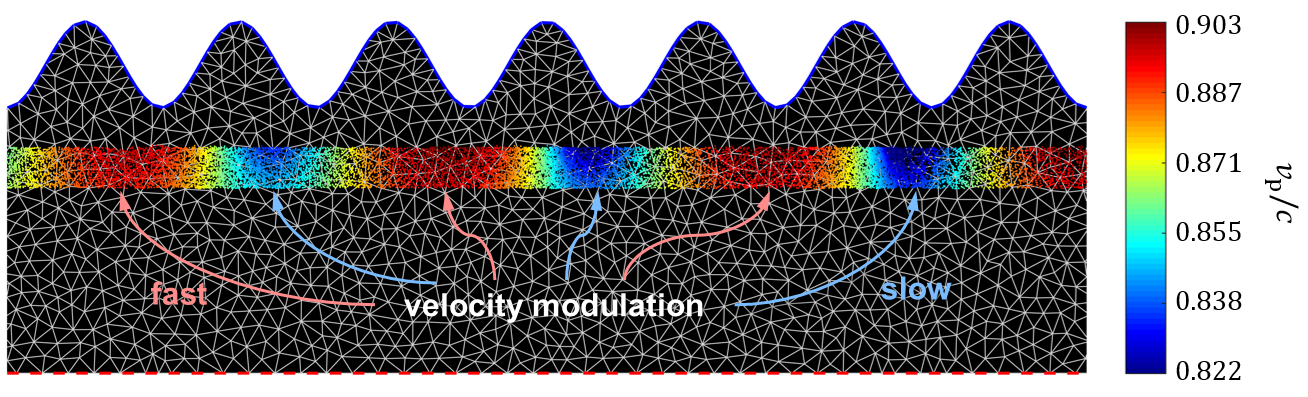}
	\caption{A zoomed-in region of four rightmost corrugations of Fig. \ref{fig:particle_distribution} with RGB color scales reflecting particle velocities.}
	\label{fig:velocity_modulation}
\end{figure}

\subsubsection{System performance}\label{sec:System performance}
Fig. \ref{fig:particle_distribution} illustrates a snapshot of the electron beam at $t=21.50$~ns.
In Fig. \ref{fig:velocity_modulation} we plot a zoomed-in beam picture that shows the four rightmost corrugations, where RGB colors indicate variations in normalized particle velocities $v_{\text{p}}/c$. The periodic particle beam bunching is a result of particles being accelerated or decelerated, which means that the beam electrons synchronously lose and recover their kinetic energy. The net energy is transferred from the beam to the waves as seen in Fig. \ref{fig:phase_plot}, which illustrates a phase space distribution of the particle beam at $24.00$~ns. 
The particles decelerated from the initial velocity ($0.877c$~m/s) dominate the accelerated particles, so that a net loss of the beam kinetic energy leads to amplification of the $\text{TM}_{01}$ mode.
\begin{figure}[t]
\centering
\includegraphics[width=2.5in]{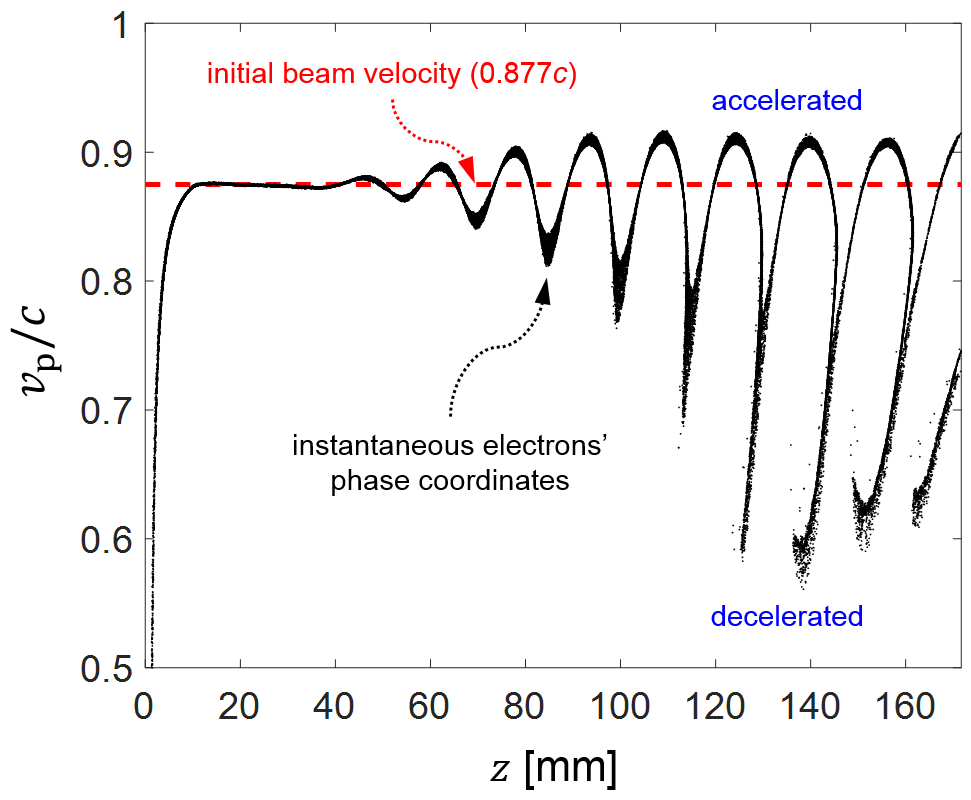}
\caption{Phase-space plot at $24.00$ ns.}
\label{fig:phase_plot}
\end{figure}
Fig. \ref{fig:self_electric_fields} presents a vector plot of self-fields generated by the electron beam at $t=76.00$~ns (in steady state).
Coherent Cerenkov beam-wave interactions give rise to a strong $\text{TM}_{01}$ mode that may be observed within the SWS region. 
\begin{figure}[t]
\centering
\includegraphics[width=4.5in]{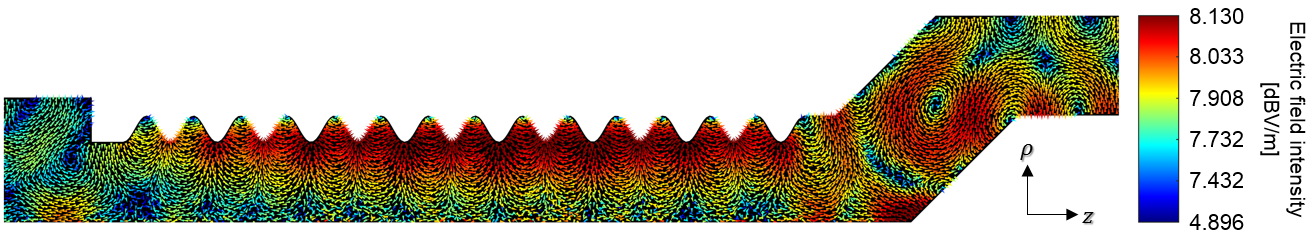}
\caption{A snapshot of steady-state self-fields ($76.00$~ns).}
\label{fig:self_electric_fields}
\end{figure}
The self-fields at the output port were analyzed in time and frequency domains, as shown in Fig. \ref{fig:output_signal_evolution} and Fig. \ref{fig:output_signal_spectrum}, respectively.
It can be seen in Fig. \ref{fig:output_signal_evolution} that although initially there are no oscillations, the output signal starts to oscillate simultaneously with the onset of beam bunching. This signal keeps evolving and eventually approaches a steady state at about $35$~ns.
By performing the Fourier analysis of steady-state output signals, we show that their spectrum shows a good degree of single-mode purity at $f_{\text{osc}}=15.575$~GHz, as shown in Fig. \ref{fig:output_signal_spectrum}.
\begin{figure}[h]
\centering
\subfloat[\label{fig:output_signal_evolution}]{
	\includegraphics[width=2.25in]{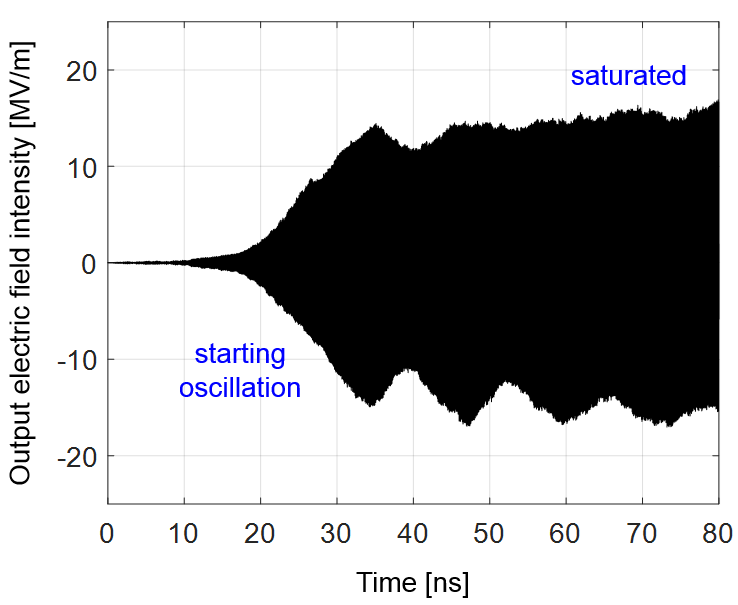}
}
\subfloat[\label{fig:output_signal_spectrum}]{
	\includegraphics[width=2.3in]{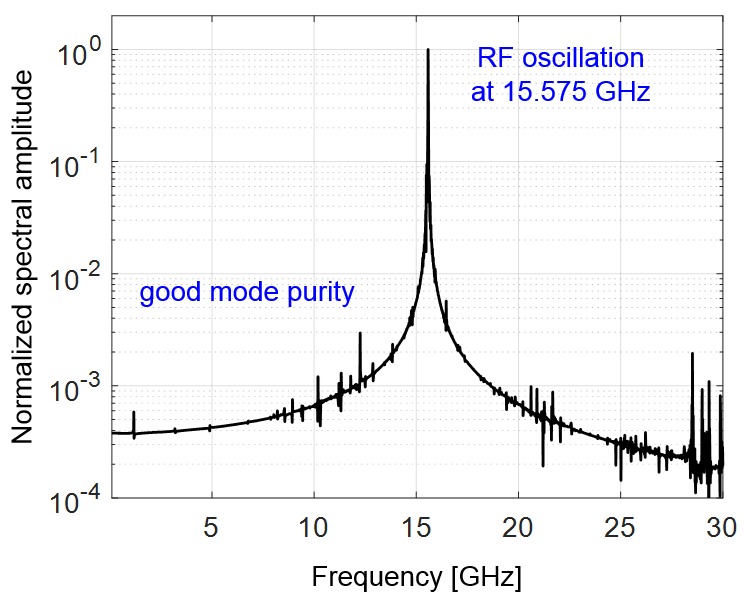}
}
\caption{Output signal analysis in (a) time and (b) frequency domains.}
\end{figure}
\begin{figure}[t!]
\centering
\subfloat[\label{fig:NR_DCE}]{
	\includegraphics[width=2.25in]{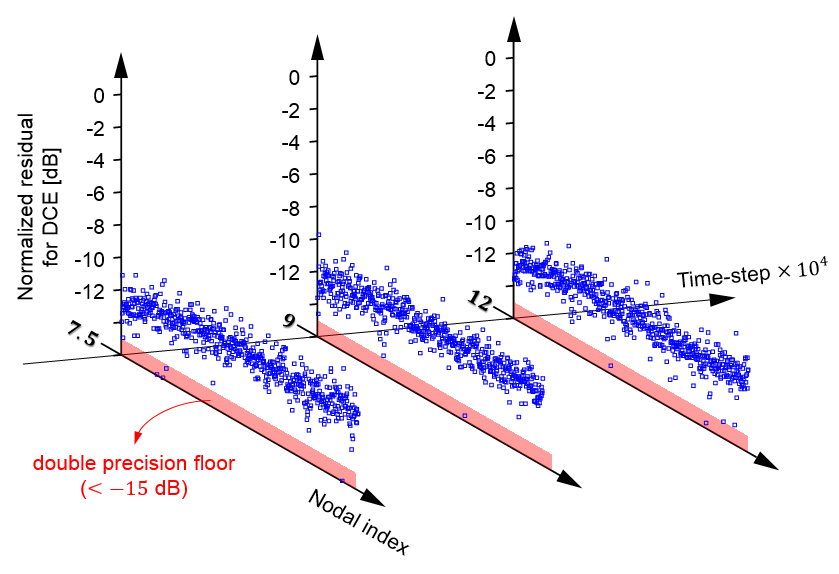}
}
\subfloat[\label{fig:NR_DGL}]{
	\includegraphics[width=2.25in]{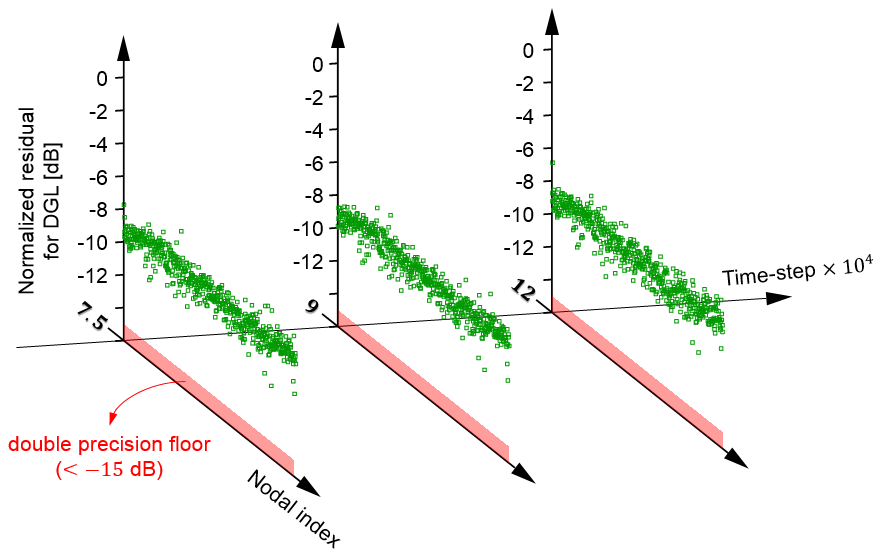}
}
\caption{Verification of charge conservation at nodes along time (at time-steps of $7.5\times10^{4}$, $9\times10^{4}$, $12\times10^{4}$) by testing NR levels of (a) DCE and (b) DGL.}
\end{figure}
In order to verify charge conservation, we plot the normalized residuals (NR) versus the nodal indices for the DCE and DGL~\cite{na2016local} in Figs. \ref{fig:NR_DCE} and Fig. \ref{fig:NR_DGL}, respectively.
These residuals, evaluated for each time step $n+\frac{1}{2}$ or $n$ and node $i$, are defined as
\begin{flalign}
{\text{NR}_{\text{DCE}}}^{n+\frac{1}{2}}_{i}
=&
1 + \frac{1}{\Delta t}
\frac{\left(\left[\mathbb{Q}\right]^{n+1}_{i}-\left[\mathbb{Q}\right]^{n}_{i}\right)}{ \left(\left[\tilde{\mathcal{D}}_{\text{div}}\right]\cdot{\left[\mathbb{J}\right]}^{n+\frac{1}{2}}\right)_{i}  },
\\
{\text{NR}_{\text{DGL}}}_{i}^{n}
=&
1 - \frac{\left[\mathbb{Q}\right]^{n}_{i}}
{ \left(
\left[\tilde{\mathcal{D}}_{\text{div}}\right]\cdot{\left[\star_{\epsilon}\right]}\cdot\left[\mathbb{E}\right]^{n}\right)_{i}}.
\end{flalign}

Fig. \ref{fig:NR_DCE} shows that NR levels for DCE remain fairly low at all nodes and very close to the double precision floor (below $10^{-15}$). Sparsely observed peaks are due to the fact that the electron-beam edges, where electrons occasionally travel through extremely small cell fractions, generate numerical noise during the scatter step. However, these errors still remain well below $10^{-10}$.
NR levels for DGL are also distributed around the double precision floor as shown in Fig. \ref{fig:NR_DGL}, which means that charge conservation is maintained within round-off errors.

\begin{figure}[t]
\centering
\includegraphics[width=4.5in]{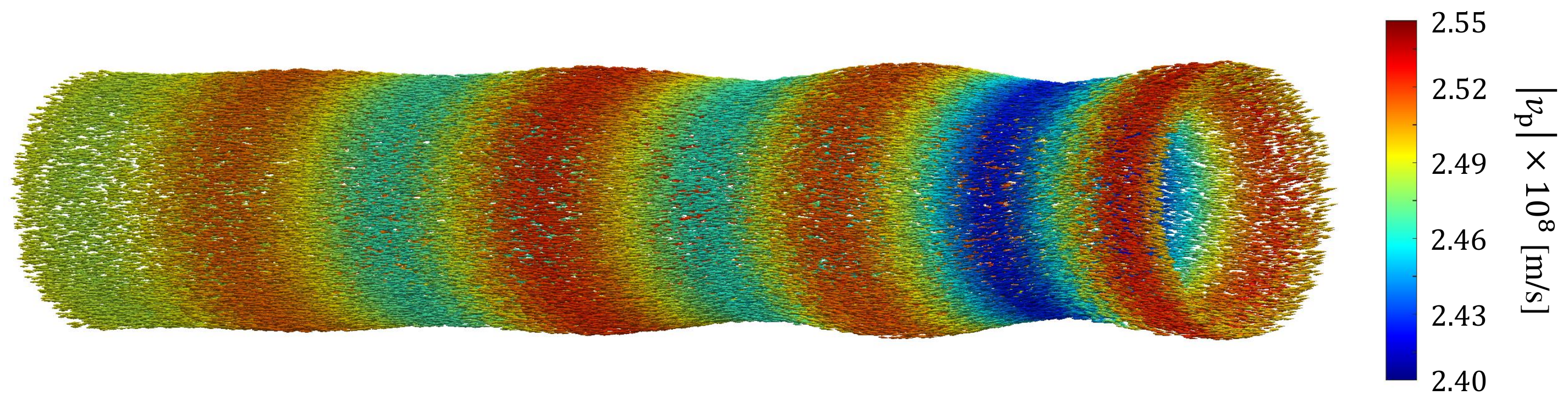}
\caption{3D velocity plots for an electron beam with the BFS magnetic field of $0.5$~T.}
\label{fig:weak_BFS}
\end{figure}
The transverse dynamics of an electron beam in the cross-section plane are sensitive to gyroradius value $\rho_{g}$ which depends on the BFS magnetic field strength. Fig. \ref{fig:weak_BFS} shows 3D electron velocity plots for a weaker BFS magnetic field reduced from $5$~T to $0.5$~T.
RGB colors again indicate particle velocity magnitudes. The weaker axial magnetic field yields a relatively larger $\rho_{g}$ in the polar ($\rho\phi$) plane, which makes the electron beam to gradually expand radially.

{\color{black}
\subsubsection{Staircasing error analysis}\label{sec:The staircasing error effect}
We now examine staircasing errors resulting from a discrete approximation of the SCSWS boundary.
We consider eight cases comprising the same BWO geometry modeled by: ($i$) a coarse mesh, ($ii$) a double-refined mesh, ($iii$) a quadruple-refined mesh, ($iv$) a octuple-refined mesh, ($v$) a coarse mesh with staircased boundary, ($vi$) a double-refined mesh with staircased boundary, ($vii$) a quadruple-refined mesh with staircased boundary, and ($viii$) a octuple-refined mesh with staircased boundary. 
Fig. \ref{fig:SCSWS_bd} depicts one period of the SCSWS boundary as rendered by the different meshes.
\begin{figure}
\centering
\includegraphics[width=4.5in]{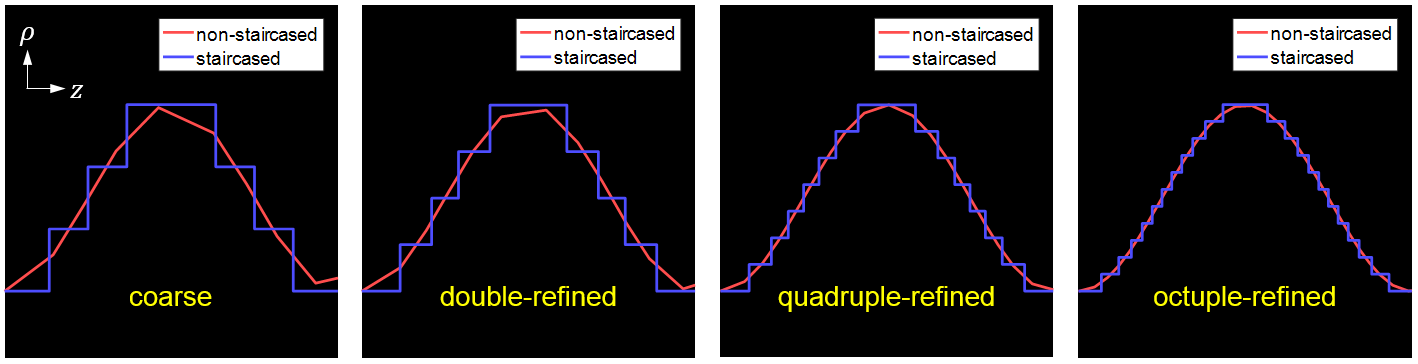}
\caption{SCSWS boundary profiles for all cases.}
\label{fig:SCSWS_bd}
\end{figure}
All cases employ unstructured meshes. 
Although unstructured meshes do not necessitate staircased boundaries, the latter are enforced to reproduce the staircasing that would be present on structured meshes. At the same time, this setup allows for a detailed study that effectively isolates staircasing error from grid-dispersion errors.
The mesh information for all cases is shown in Table~\ref{tab:mesh_info}.
\begin{table}[t]
\caption{Mesh information for different SCSWS cases} 
\centering
{\scriptsize
\begin{tabular}{c c c c c}
\hline\hline
\\[-0.5em]
Case & $N_{0}$ & $N_{1}$ & $N_{2}$ & ave$\left(l_{\text{edge}}\right)$ [mm] \\
\\[-0.5em]
\hline
\\[-0.75em]
coarse & 4,564 & 12,978 & 8,415 & 2.50\\
double-refined & 5,801 & 16,545 & 10,745 & 1.67 \\
quadruple-refined & 8,771 & 25,236 & 16,466 & 1.11 \\
octuple-refined & 15,333 & 44,542 & 29,210 & 0.74 \\
coarse staircased & 4,593 & 13,005 & 8,413 & 2.50 \\
double-refined staircased & 5,909 & 16,766 & 10,858 & 1.67 \\
quadruple-refined staircased & 8,939 & 25,547 & 16,610 & 1.11 \\
octuple-refined staircased & 15,543 & 44,904 & 29,362 & 0.74 \\
\\[-0.75em]
\hline
\end{tabular}
}
\label{tab:mesh_info}
\end{table}
We have performed hot tests for all cases, keeping the number of superparticle per each cell by $106$ on average.
\begin{figure}[t]
\centering
\subfloat[\label{fig:stea}]{\includegraphics[width=2.2in]{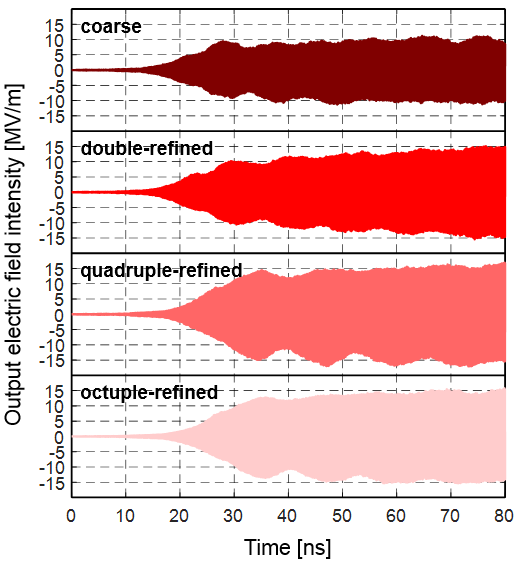}}
~~
\subfloat[\label{fig:steb}]{\includegraphics[width=2.2in]{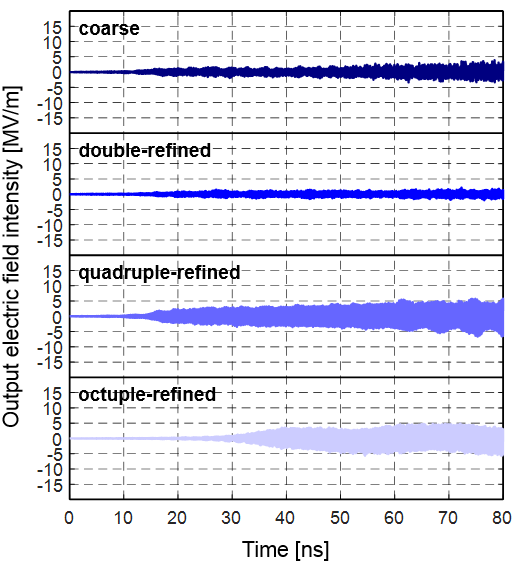}}
\caption{Field signal at the output port in (a) SCSWS and (b) staircased SCSWS in the time domain. 
}
\label{fig:evelope_field_evolution_comp}
\end{figure}

\begin{figure}[t]
\centering
\includegraphics[width=3in]{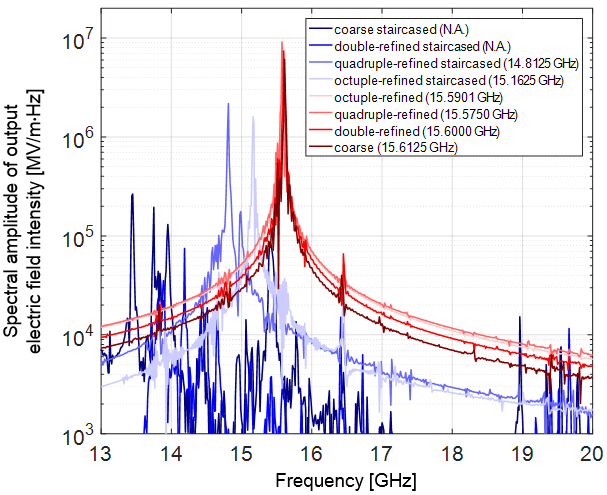}
\caption{Normalized spectral amplitude at the output port in SCSWS and staircased SCSWS. 
}
\label{fig:spectrum_comp}
\end{figure}
Field output signals in time and oscillation frequencies are displayed in Fig. \ref{fig:evelope_field_evolution_comp} and Fig.~\ref{fig:spectrum_comp}.
The results based on meshes devoid of staircasing error presented in  Fig.~\ref{fig:stea} 
converge much faster than those of meshes with staircased boundaries in  Fig.~\ref{fig:steb} 
The oscillation frequency values in Fig. \ref{fig:spectrum_comp} also shows the fast convergent rate of present EM-PIC simulations.
In particular, it is interesting to see that the results from the BWO with a staircased SCSWS are unable to capture much RF oscillation at all in the case of coarse and double-refined meshes (case (v) and (vi)).
The RF oscillation is more visible on quadruple- or octuple-refined meshes (case (vii) and (viii)).
This is because underestimation of the $\pi$-point frequency\footnote{$\pi$-point denotes the solution of the modal field which have maximum frequency in the passband.} in the modal dispersion causes a slow-charge mode driven by the electron beam~\cite{benford2015high} to falls into the forward wave region and, as a result, the system does not act as an oscillator anymore.
Note that the BWO system here is designed to operate at $\pi$-point of the modal fields.
The underestimation of the $\pi$-point frequency in the staircased boundary is shown by Fig. \ref{fig:dispersion_comp}, which depicts the dispersion diagrams for the $\text{TM}_{01}$ mode of each SWS. The analytic dispersion relation for the $\text{TM}_{01}$ mode is obtained based on Floquet's theory by considering harmonics up to $6^{\text{th}}$ order~\cite{barroso2002cylindrical}.
It is clearly seen that the $\pi$-point frequency decreases as the staircasing error become significant. On the other hand, the cases without staircasing errors rapidly converge to the analytic prediction.   

\begin{figure}[t]
\centering
\includegraphics[width=3in]{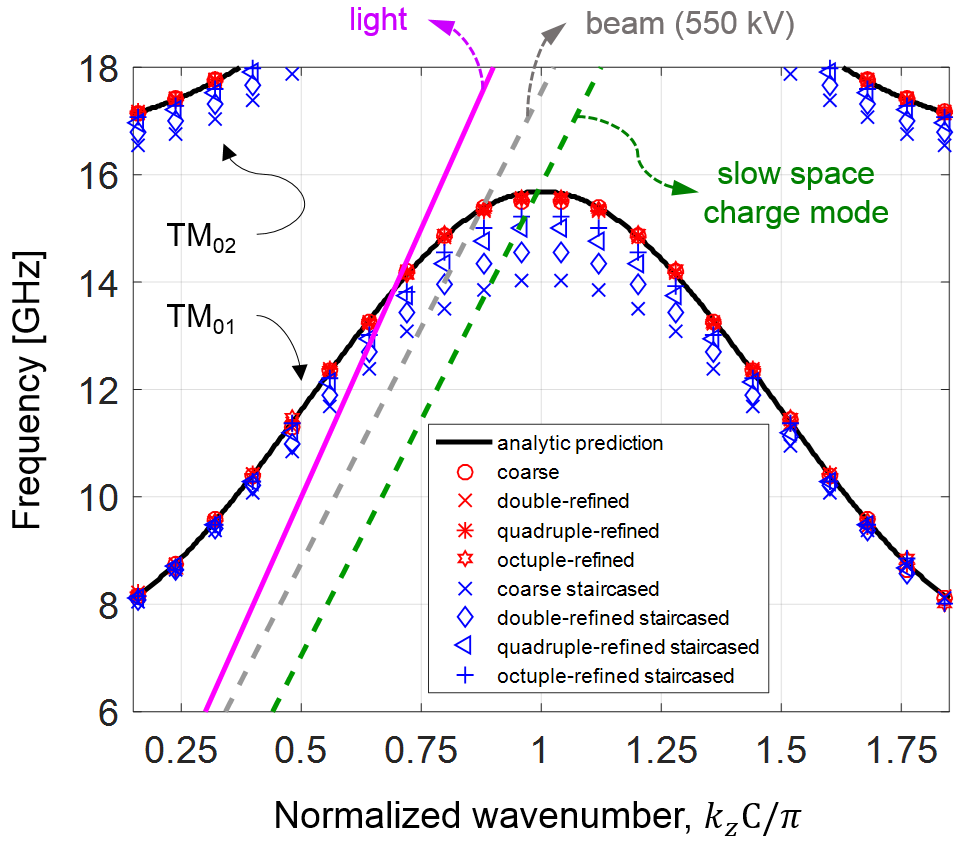}
\caption{Dispersion relations from ``cold tests''.}
\label{fig:dispersion_comp}
\end{figure}
}

{\color{black}
\section{Concluding Remarks}\label{sec:Conclusion}
We introduced a new axisymmetric charge-conservative EM-PIC algorithm on unstructured grids for the analysis and design of micromachined VEDs with cylindrical axisymmetry. 
We demonstrated that cylindrical symmetry in device geometry, fields, and sources enables reduction of the 3D problem to a 2D one through  considering $\text{TE}^{\phi}$-polarized fields in the meridian ($\rho z$) plane and  introducing an artificial inhomogeneous medium to account for the metric elements of the cylindrical coordinate system.
As a result, computational resources are significantly reduced. 
The unstructured-grid spatial discretization is achieved by using Whitney forms in the meridian plane. 
Using leapfrog time integration, we obtain space- and time-discretized Maxwell's equations in the form of a mixed $\mathcal{E}-\mathcal{B}$ FETD scheme.
A local and explicit time update is made possible by employing the SPAI approach~\cite{na2016local}.
Interpolation of the field values to the particles' positions is also performed by Whitney forms and used for solving the Newton-Lorentz equations of motion of each particle.
Relativistic particles are accelerated based on a relativistic Boris algorithm with correction.
In the particle scatter step, we employ a recently introduced Cartesian charge-conserving scatter scheme for unstructured grids~\cite{moon2015exact}. The algorithm was validated by considering cylindrical cavity and space-charge-limited cylindrical diode problems.  We also illustrated the advantages of the present algorithm in the analysis of a  BWO system including a slow-wave waveguide structure with complex geometry.

\section*{Acknowledgment}
This work was supported in part by U.S. NSF grant ECCS-1305838, OSC grants PAS-0061 and PAS-0110, FAPESP-OSU grant 2015/50268-5. H. Moon and D.-Y. Na were also supported by
the OSU Presidential Fellowship program. 
}
\clearpage
\appendix
{\color{black}
\section{Pairing operation and generalized Stokes theorem}\label{sec:app_A}
One of the key properties of Whitney $p$-forms is that they admit a natural ``pairing'' with the $p$-cells of the mesh, where $p$ refers to the dimensionality i.e. $p=0$ refers to nodes, $p=1$ to edges, and so on~\cite{teixeira1999lattice}.  Computationally, the pairing operation between
  an $i$-th $p$-cell of the grid $\sigma_{p}^{i}$ and 
the Whitney form $w^{p}_{j}$ associated with the $j$-th $p$-cell is effected by the integral below and yields~\cite{teixeira1999lattice,teixeira2014lattice}
\begin{eqnarray}
\left<\sigma_{p}^{i},w_{j}^{p}\right>=\int_{\sigma_{p}^{i}} w^{p}_{j}=\delta_{i,j},
\label{eq:paring}
\end{eqnarray}
for $p=0,1,2,3$ in the 3D space and where $\delta_{i,j}$ is the Kronecker delta.
The generalized Stokes theorem~\cite{teixeira1999lattice,teixeira2014lattice,cairns1936the} states
\begin{eqnarray}
\left<\sigma_{p+1},d w^{p}_{j}\right>
=
\left<\left(\partial \sigma_{p+1}\right)_{p},w^{p}_{j}\right>
\label{eq:GST}
\end{eqnarray}
where $\partial$ is the boundary operator.
Using the above pairing operation and generalized Stokes' theorem, we can obtain discrete Maxwell's equation on a irregular lattice (unstructured grid).
For example, applying pairing for $2$-cells into Faraday's law on the primal mesh gives
\begin{eqnarray}
\left<\sigma^{i}_{2},d\mathcal{E}\right>
\!\!\!\!&=&\!\!\!\!
\left<\sigma^{i}_{2},\frac{\partial \mathcal{B}}{\partial t}\right>,
\label{eq:paring_GST_FL_1}
\end{eqnarray}
and applying generalized Stokes' theorem into the left-hand side term of (\ref{eq:paring_GST_FL_1}) yields 
\begin{eqnarray}
\left<\left(\partial \sigma^{i}_{2}\right)_{1},\mathcal{E}\right>
\!\!\!\!&=&\!\!\!\!
\frac{\partial }{\partial t}\left<\sigma^{i}_{2},\mathcal{B}\right>.
\label{eq:paring_GST_FL_2}
\end{eqnarray}
Substituting (\ref{eq:E_expr}) and (\ref{eq:B_expr}) into (\ref{eq:paring_GST_FL_2}) and using $\left(\partial \sigma^{i}_{2}\right)_{1}=\sum_{k=1}^{N_{1}}C_{ik}\sigma^{k}_{1}$ where $C_{ik}$ is an incidence matrix element taking a value in the set of $\left\{-1,0,1\right\}$~\cite{teixeira1999lattice,teixeira2014lattice,hughes1981lagrangian,guth1980existence}, we obtain
\begin{eqnarray}
\left<\sum_{k=1}^{N_{1}}C_{ik} \sigma^{k}_{1}, \sum_{j=1}^{N_{1}}\mathbb{E}_{j}w_{j}^{1}\right>
\!\!\!\!&=&\!\!\!\!
\frac{\partial }{\partial t}\left<\sigma^{i}_{2},\sum_{j=1}^{N_{2}}\mathbb{B}_{j}w_{j}^{2}\right>.
\label{eq:paring_GST_FL_3}
\end{eqnarray}
By using (\ref{eq:paring}), (\ref{eq:paring_GST_FL_3}) can be rewritten as
\begin{eqnarray}
\sum_{k=1}^{N_{1}}C_{ik} \mathbb{E}_{k}
\!\!\!\!&=&\!\!\!\!
\frac{\partial }{\partial t}\mathbb{B}_{i},
\label{eq:paring_GST_FL_4}
\end{eqnarray}
for $i=1,...,N_{2}$.
(\ref{eq:paring_GST_FL_4}) represents the discrete representation of Faraday's law as written in (\ref{eq:DFL_mat}).
Discrete Ampere's law can be obtained by a similar procedure on the dual mesh.

\clearpage
\section{Discrete Hodge matrix}\label{sec:app_B}
Using vector calculus proxies~\cite{he2007differential,kim2011parallel,tarhasaari1999some,gillette2011dual,he2006geometric} and considering a (doubly) artificial inhomogeneous medium at the meridian plane as done in (\ref{eq:HS_H_RHS}) and (\ref{eq:HS_D_RHS}), we can rewrite (\ref{eq:dis_hodge_eps}) and (\ref{eq:dis_hodge_mu}) as
\begin{eqnarray}
\left[\star_{\epsilon}\right]_{i,j}
\!\!\!\!&=&\!\!\!\!
\int_{\Omega}\left(\epsilon_{0}\rho\right) \vec{W}_{i}^{1}\left(z,\rho\right)\cdot\vec{W}_{j}^{1}\left(z,\rho\right) dz d\rho,
\\
\left[\star_{\mu}^{-1}\right]_{i,j}
\!\!\!\!&=&\!\!\!\!
\int_{\Omega}\left(\mu^{-1}_{0}\rho\right) \vec{W}_{i}^{2}\left(z,\rho\right)\cdot\vec{W}_{j}^{2}\left(z,\rho\right) dz d\rho,
\end{eqnarray}
where $\vec{W}_{i}^{p}$ is a vector proxy for Whitney $p$-form associated with the $i$-th $p$-cell.
The vector proxies for Whitney 1- and 2-forms are written as \cite{moon2015exact}
\begin{eqnarray}
\vec{W}_{i}^{1}
\!\!\!\!&=&\!\!\!\!
\lambda_{i_{a}}\vec{\nabla}\lambda_{i_{b}}-\lambda_{i_{b}}\vec{\nabla} \lambda_{i_{a}}
\\
\vec{W}_{i}^{2}
\!\!\!\!&=&\!\!\!\!
2\times
\left[
\lambda_{i_{a}}\vec{\nabla}\lambda_{i_{b}}\times\vec{\nabla} \lambda_{i_{c}}+
\lambda_{i_{b}}\vec{\nabla}\lambda_{i_{c}}\times\vec{\nabla} \lambda_{i_{a}}+
\lambda_{i_{c}}\vec{\nabla}\lambda_{i_{a}}\times\vec{\nabla} \lambda_{i_{b}}
\right]
\end{eqnarray}
where $i_{a}$, $i_{b}$, and $i_{c}$ denote indices for grid nodes belonging to the $i$-th $p$-cell, for $p=1$ or $2$, and
$\lambda$ denote the barycentric coordinates of each node. 
Since Whitney forms have compact support, we can express the global discrete Hodge matrix in terms of sum of local matrices for the $i$-th triangle as
\begin{eqnarray}
\left[\mathcal{T}\right]_{i}=\Delta_{i}
\begin{bmatrix}
T_{1,1} & T_{1,2} & T_{1,3} \\ T_{2,1} & T_{2,2} & T_{2,3} \\ T_{3,1} & T_{3,2} & T_{3,3}
\end{bmatrix}
\end{eqnarray}
where $\Delta_{i}$ is the area of $i$-th triangle, $t_{a}=\frac{\vec{\nabla}\lambda_{1}\cdot\vec{\nabla}\lambda_{1}}{5}$, $t_{b}=\frac{\vec{\nabla}\lambda_{1}\cdot\vec{\nabla}\lambda_{2}}{5}$, $t_{c}=\frac{\vec{\nabla}\lambda_{2}\cdot\vec{\nabla}\lambda_{2}}{5}$, and $\rho_{1}$, $\rho_{2}$, $\rho_{3}$ denote the $\rho$ coordinate in polar coordinate systems with origin at the grid nodes belonging to the $i$-th $2$-cell (triangle), and
\begin{eqnarray}
&& \!\!\!\!\!\!\!\!\!\!\!\!\!\!\!\!\!\!\!\!\!\!\!\!\!\!\!\!
T_{1,1}
=
t_{a}
\left(\frac{\rho_{1}}{6}+\frac{\rho_{2}}{2}+\frac{\rho_{3}}{6}\right)
-
t_{b}
\left(\frac{\rho_{1}}{3}+\frac{\rho_{2}}{3}+\frac{\rho_{3}}{6}\right)
+
t_{c}
\left(\frac{\rho_{1}}{2}+\frac{\rho_{2}}{6}+\frac{\rho_{3}}{6}\right),
\\
&& \!\!\!\!\!\!\!\!\!\!\!\!\!\!\!\!\!\!\!\!\!\!\!\!\!\!\!\!
T_{1,2}
=
t_{a}
\left(\frac{\rho_{1}}{4}+\frac{\rho_{2}}{3}+\frac{\rho_{3}}{4}\right)
-
t_{b}
\left(\frac{\rho_{1}}{2}+\frac{\rho_{2}}{12}+\frac{\rho_{3}}{4}\right)
-t_{c}
\left(\frac{\rho_{1}}{2}+\frac{\rho_{2}}{6}+\frac{\rho_{3}}{6}\right),
\\
&& \!\!\!\!\!\!\!\!\!\!\!\!\!\!\!\!\!\!\!\!\!\!\!\!\!\!\!\!
T_{1,3}
=
t_{a}
\left(\frac{\rho_{1}}{6}+\frac{\rho_{2}}{2}+\frac{\rho_{3}}{6}\right)
+
t_{b}
\left(\frac{\rho_{1}}{12}+\frac{\rho_{2}}{2}+\frac{\rho_{3}}{4}\right)
-t_{c}
\left(\frac{\rho_{1}}{3}+\frac{\rho_{2}}{4}+\frac{\rho_{3}}{4}\right),
\\
&& \!\!\!\!\!\!\!\!\!\!\!\!\!\!\!\!\!\!\!\!\!\!\!\!\!\!\!\!
T_{2,1}=T_{1,2},
\\
&& \!\!\!\!\!\!\!\!\!\!\!\!\!\!\!\!\!\!\!\!\!\!\!\!\!\!\!\!
T_{2,2}
=
t_{a}
\left(\rho_{1}+\frac{\rho_{2}}{2}+\rho_{3}\right)
+
t_{b}
\left(\frac{\rho_{1}}{3}+\frac{\rho_{2}}{2}+\frac{\rho_{3}}{3}\right)
+
t_{c}
\left(\frac{\rho_{1}}{2}+\frac{\rho_{2}}{6}+\frac{\rho_{3}}{6}\right),
\\
&& \!\!\!\!\!\!\!\!\!\!\!\!\!\!\!\!\!\!\!\!\!\!\!\!\!\!\!\!
T_{2,3}=
t_{a}
\left(\frac{\rho_{1}}{4}+\frac{\rho_{2}}{3}+\frac{\rho_{3}}{4}\right)
+
3t_{b}
\left(\frac{\rho_{1}}{4}+\frac{\rho_{2}}{4}+\frac{\rho_{3}}{3}\right)
+
t_{c}
\left(\frac{\rho_{1}}{3}+\frac{\rho_{2}}{4}+\frac{\rho_{3}}{4}\right),
\\
&& \!\!\!\!\!\!\!\!\!\!\!\!\!\!\!\!\!\!\!\!\!\!\!\!\!\!\!\!
T_{3,1}=T_{1,3},
\\
&& \!\!\!\!\!\!\!\!\!\!\!\!\!\!\!\!\!\!\!\!\!\!\!\!\!\!\!\!
T_{3,2}=T_{2,3},
\\
&& \!\!\!\!\!\!\!\!\!\!\!\!\!\!\!\!\!\!\!\!\!\!\!\!\!\!\!\!
T_{3,3}=
t_{a}
\left(\frac{\rho_{1}}{6}+\frac{\rho_{2}}{2}+\frac{\rho_{3}}{6}\right)
+
4t_{b}
\left(\frac{\rho_{1}}{8}+\frac{\rho_{2}}{3}+\frac{\rho_{3}}{6}\right)
+
t_{c}
\left(\frac{\rho_{1}}{2}+\rho_{2}+\rho_{3}\right).
\end{eqnarray}
The elements of $\left[\star_{{\mu}^{-1}}\right]$ in 2D take the form of
\begin{eqnarray}
\left[\star_{{\mu}^{-1}}\right]_{i,j}=\delta_{i,j}\frac{4\Delta_{j}\sum_{n=1}^{3}\rho_{n}}{3}\vec{\nabla}\lambda_{1}\times\vec{\nabla}\lambda_{2}.
\end{eqnarray}
The sparsity patterns for discrete Hodge matrices of the mesh exemplified in Fig. \ref{fig:mesh_example} are provided in Fig. \ref{fig:sp_DHM}.
Due to finite support of Whitney forms, these matrices are sparse. 
In the 2-D case, each row of the permittivity Hodge matrix has only 5 nonzero elements (except for the rows associated to boundary edges) and the permeability Hodge matrix is a diagonal matrix.
\begin{figure}
\centering
\subfloat[\label{fig:eps_mat}]{\includegraphics[width=2.2in]{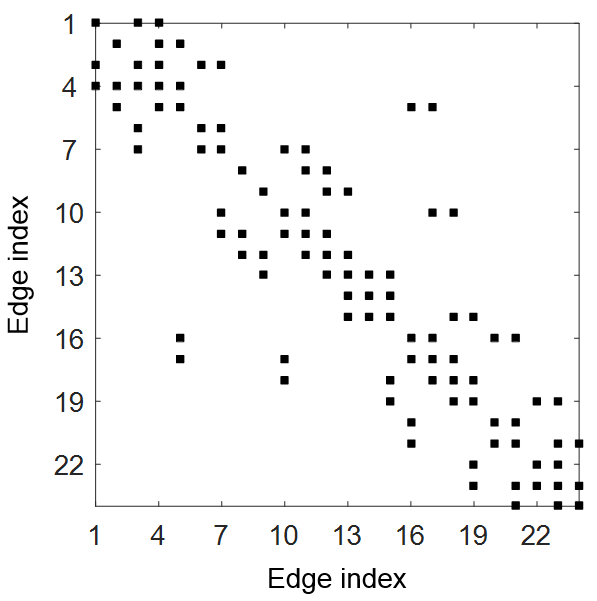}}
\subfloat[\label{fig:mu_mat}]{\includegraphics[width=2.2in]{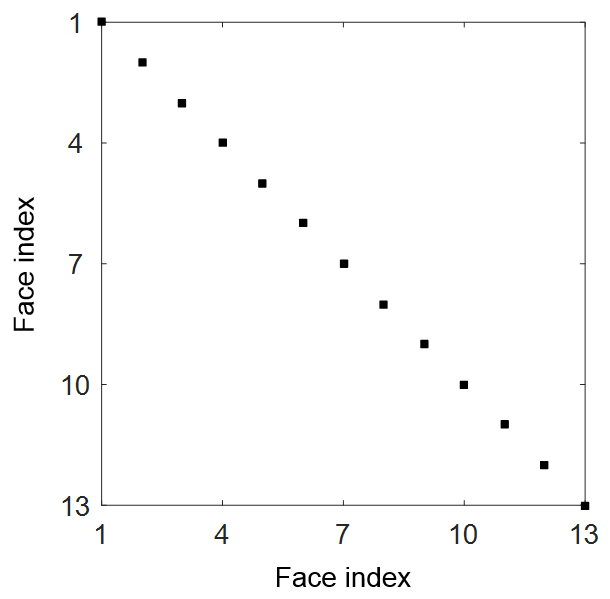}}
\caption{Sparsity patterns for discrete Hodge matrices of (a) $\left[\star_{\epsilon}\right]$ and (b) $\left[\star_{\mu^{-1}}\right]$.}
\label{fig:sp_DHM}
\end{figure}

\clearpage
\section{Barycentric dual lattice relations}\label{sec:app_C}
The barycentric dual lattice~\cite{teixeira2014lattice,sen2000geometric} has a similar contraction identity to (\ref{eq:paring}), given by
\begin{eqnarray}
\left<\tilde{\sigma}_{n-p}^{i},\star w_{j}^{p}\right>=\int_{\tilde{\sigma}_{n-p}^{i}}\star w_{j}^{p}=\delta_{i,j}
\end{eqnarray}
where $n$ is the ambient space dimension and $p$ is the grid element dimension.
Since we consider Hodge duals of electric current and charge densities in the primal mesh, we need to express them in dual formulation to be used in discrete Ampere's law.
The electric current density can be expressed by
\begin{eqnarray}
\mathcal{J}=\star\mathcal{J}_{\star}=\star\left(\sum_{j=1}^{N_{1}}\mathbb{J}_{\star,j} w_{j}^{1}\right)=\sum_{j=1}^{N_{1}}\mathbb{J}_{\star,j} \, \star(w_{j}^{1})
\end{eqnarray}
In order to obtain the discrete representation for Ampere's law, we combine the pairing $\left<\tilde{\sigma}_{1}^{i},\left(\cdot\right)\right>$ with Ampere's law. The electric current density term become 
\begin{eqnarray}
\left<\tilde{\sigma}_{1}^{i},\sum_{j=1}^{N_{1}}\mathbb{J}_{\star,j} \star w_{j}^{1}\right>
=
\sum_{j=1}^{N_{1}}\mathbb{J}_{\star,j}\left<\tilde{\sigma}_{1}^{i},\star w_{j}^{1}\right>
=
\sum_{j=1}^{N_{1}}\delta_{i,j} \, \mathbb{J}_{\star,j}.
\end{eqnarray}
Therefore, we can write above in matrix form as
\begin{eqnarray}
\left[\mathbb{J}\right]
=
\left[\mathcal{I}\right]\cdot\left[\mathbb{J}_{\star}\right]
\end{eqnarray}
where $\left[\mathbb{J}\right]$ is a column vector with all DoFs for $\mathcal{J}$ expanded in terms of Whitney 1-forms on the dual mesh, and $\left[\mathcal{I}\right]$ is the identity matrix.
A similar procedure can be done for the electric charge density in Gauss' law:
\begin{eqnarray}
\left[\mathbb{Q}\right]
=
\left[\mathcal{I}\right]\cdot\left[\mathbb{Q}_{\star}\right]
\end{eqnarray}
where $\left[\mathbb{Q}\right]$ is a column vector with all DoFs for $\mathcal{Q}$ expanded in terms of Whitney 2-forms on the dual mesh.

\clearpage
\section{Structure of the incidence matrices}
\begin{figure}[t]
\centering
\includegraphics[width=2.5in]{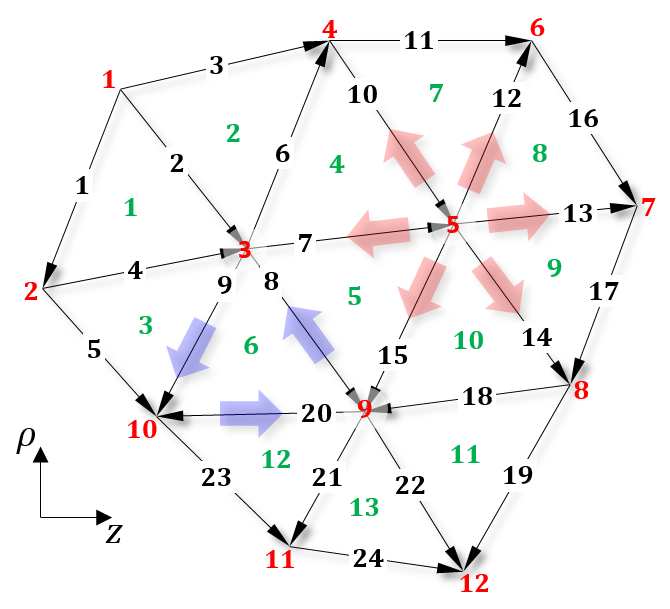}
\caption{Example (primal) unstructured mesh.}
\label{fig:mesh_example}
\end{figure}
\begin{figure}[t]
\centering
\subfloat[\label{fig:inc_mat}]{	\includegraphics[width=4.9in]{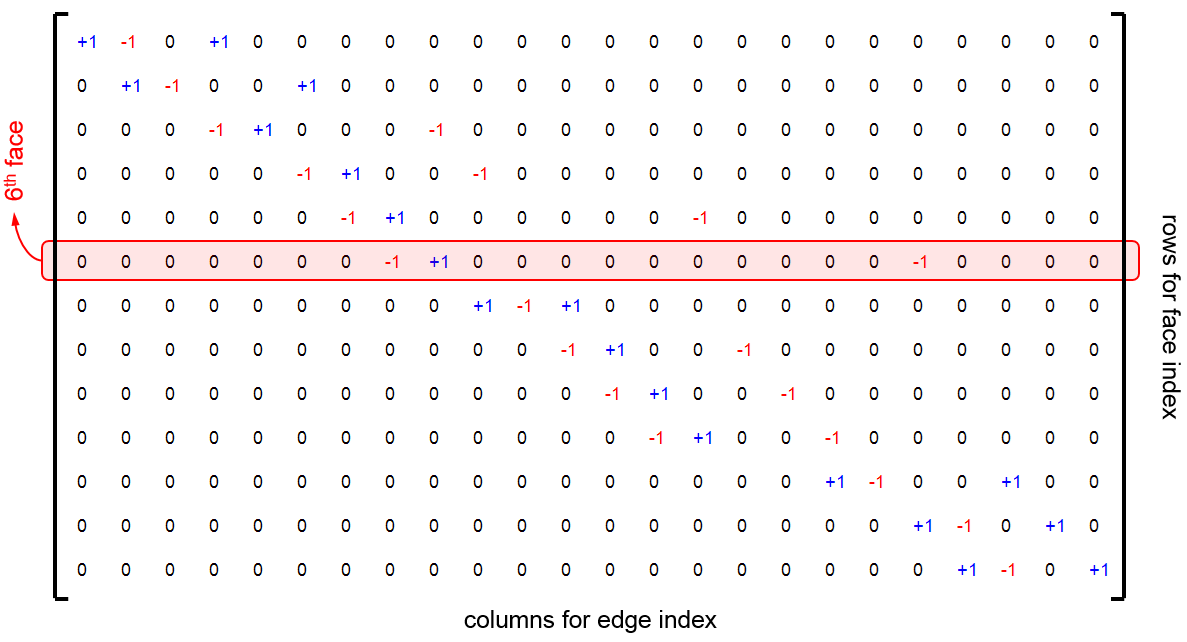}}
\\
\subfloat[\label{fig:div_mat}]{	\includegraphics[width=4.9in]{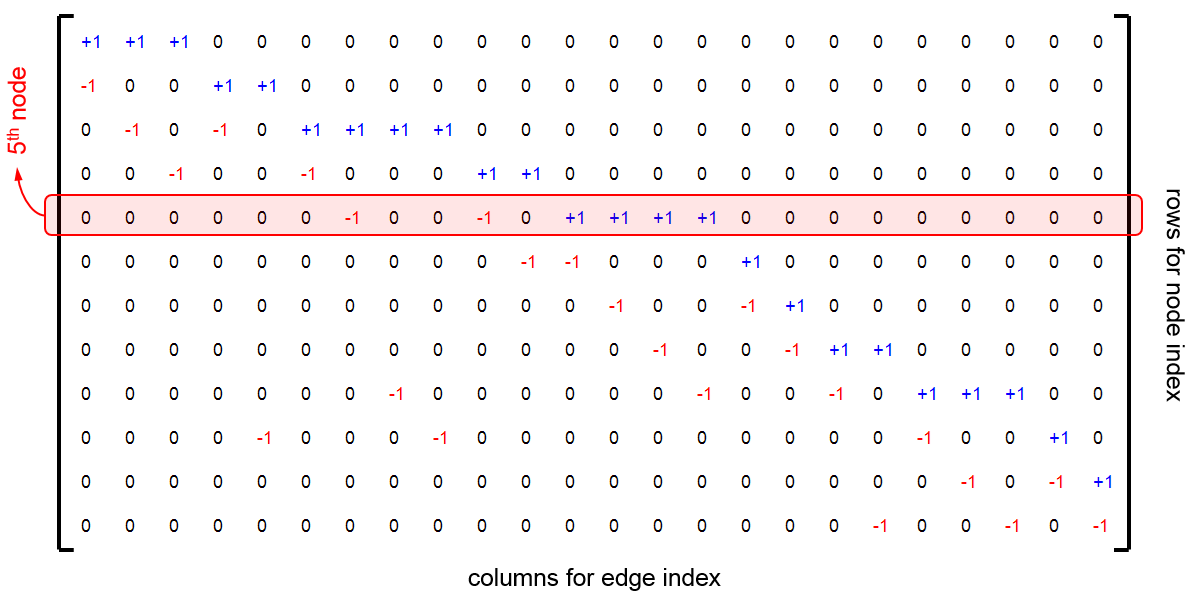}}
\caption{Incidence matrices for (a) curl $\left[\mathcal{D}_{\text{curl}}\right]$ and (b) divergence $\left[\tilde{\mathcal{D}}_{\text{div}}\right]^{T}$ operators for the mesh in Fig. \ref{fig:mesh_example}.}
\label{fig:incidence_matrices}
\end{figure}
Incidence matrices represent the spatial differential operators on the mesh (at the discrete level), distilled from the metric structure (i.e., the discrete exterior derivative)~\cite{teixeira1999lattice,teixeira2014lattice}. Since the discrete exterior derivative can be seen as the dual of the boundary operator acting on the mesh elements by the generalized Stokes theorem, Eq.~(\ref{eq:GST}), incidence matrices encode the relationship between each element of the mesh and its boundary elements (say, between an edge and its boundary nodes, between a face element and its boundary edges, and so on). 

In order to illustrate the structure of the incidence matrices, we consider a simple mesh with perfect magnetic conductor (or free edges) boundaries as depicted in Fig. \ref{fig:mesh_example}.
Red-colored numbers denote nodal indices, black-colored numbers edge indices, and green-colored number face indices. 
Intrinsic edge orientation is defined by ascending index order of the two nodes associated with any given edge.
The sign the corresponding 
$\left[\mathcal{D}_{\text{curl}}\right]$ matrix
elements is determined by comparing the intrinsic orientation of each edge with the curl and divergence operator directions indicated by the blue and beige thick arrows in Fig. \ref{fig:mesh_example}: if they are opposite, the element is $-1$, otherwise it is $+1$. For example, if we consider $\left[\mathcal{D}_{\text{curl}}\right]$, of size $N_{2} \times N_{1}$, there are three edges wrapping face number 6: edges 8, 9, and 20. As a result, $\left[\mathcal{D}_{\text{curl}}\right]_{6,8}=-1$, $\left[\mathcal{D}_{\text{curl}}\right]_{6,9}=+1$, and $\left[\mathcal{D}_{\text{curl}}\right]_{6,9}=-1$. Furthermore, $\left[\mathcal{D}_{\text{curl}}\right]_{6,k}=0$ for all other edges $k$. This is represented in Fig. \ref{fig:inc_mat}, which shows the entire  $\left[\mathcal{D}_{\text{curl}}\right]$ for this mesh.

Likewise, if we consider 
$\left[\tilde{\mathcal{D}}_{\text{div}}\right]$, of size $N_{0} \times N_{1}$, there are six edges connected to node 5: edges 7, 10, 12, 13, 14, and 15. The corresponding elements are $\left[\mathcal{D}_{\text{div}}\right]^{T}_{5,7}=-1$, $\left[\mathcal{D}_{\text{div}}\right]^{T}_{5,10}=-1$, $\left[\mathcal{D}_{\text{div}}\right]^{T}_{5,12}=+1$, $\left[\mathcal{D}_{\text{div}}\right]^{T}_{5,13}=+1$, $\left[\mathcal{D}_{\text{div}}\right]^{T}_{5,14}=+1$, and $\left[\mathcal{D}_{\text{div}}\right]^{T}_{5,15}=+1$. This is represented in Fig. \ref{fig:div_mat}.

\clearpage
\section*{References}
\bibliography{mybibfile}

\begin{thebibliography}{10}
\expandafter\ifx\csname url\endcsname\relax
  \def\url#1{\texttt{#1}}\fi
\expandafter\ifx\csname urlprefix\endcsname\relax\def\urlprefix{URL }\fi
\expandafter\ifx\csname href\endcsname\relax
  \def\href#1#2{#2} \def\path#1{#1}\fi

\bibitem{gold1997review}
S.~H. Gold, G.~S. Nusinovich, Review of high-power microwave source research,
  Rev. Sci. Instrum. 68 (1997) 3945--3974.
\newblock \href {http://dx.doi.org/http://dx.doi.org/10.1063/1.1148382}
  {\path{doi:http://dx.doi.org/10.1063/1.1148382}}.

\bibitem{cairns1997generation}
R.~A. Cairns, A.~D.~R. Phelps, Generation and Application of High Power
  Microwaves, CRC Press, New York, 1997.

\bibitem{booske2011vacuum}
J.~H. Booske, R.~J. Dobbs, C.~D. Joye, C.~L. Kory, G.~R. Neil, G.-S. Park,
  J.~Park, R.~J. Temkin, Vacuum electronic high power terahertz sources, IEEE
  Trans. THz Sci. Technol. 1 (2011) 54--75.
\newblock \href {http://dx.doi.org/10.1109/TTHZ.2011.2151610}
  {\path{doi:10.1109/TTHZ.2011.2151610}}.

\bibitem{booske2008plasma}
J.~H. Booske, Plasma physics and related challenges of
  millimeter-wave-to-terahertz and high power microwave generation, Phys.
  Plasmas 15 (2008) 055502.
\newblock \href {http://dx.doi.org/http://dx.doi.org/10.1063/1.2838240}
  {\path{doi:http://dx.doi.org/10.1063/1.2838240}}.

\bibitem{li2013experimental}
X.~Li, J.~Wang, J.~Sun, Z.~Song, H.~Ye, Y.~Zhang, L.~Zhang, L.~Zhang,
  Experimental study on a high-power subterahertz source generated by an
  overmoded surface wave oscillator with fast startup, IEEE Trans. Electron
  Devices 60 (2013) 2931--2935.
\newblock \href {http://dx.doi.org/10.1109/TED.2013.2273489}
  {\path{doi:10.1109/TED.2013.2273489}}.

\bibitem{gaponov1994applications}
A.~V. Gaponov-Grekhov, V.~L. Granatstein, Applications of High-Power
  Microwaves, Artech House, Norwood, 1994.

\bibitem{schamiloglu2004high}
E.~Schamiloglu, High power microwave sources and applications, in: 2004 IEEE
  MTT-S International Microwave Symposium Digest (IEEE Cat. No.04CH37535),
  Vol.~2, 2004, pp. 1001--1004.
\newblock \href {http://dx.doi.org/10.1109/MWSYM.2004.1339150}
  {\path{doi:10.1109/MWSYM.2004.1339150}}.

\bibitem{shiffler1990high}
D.~Shiffler, J.~A. Nation, G.~S. Kerslick, A high-power, traveling wave tube
  amplifier, IEEE Trans. Plasma Sci. 18 (1990) 546--552.
\newblock \href {http://dx.doi.org/10.1109/27.55926}
  {\path{doi:10.1109/27.55926}}.

\bibitem{johnson1955backward}
H.~R. Johnson, Backward-wave oscillators, Proc. IRE 43 (1955) 684--697.
\newblock \href {http://dx.doi.org/10.1109/JRPROC.1955.278054}
  {\path{doi:10.1109/JRPROC.1955.278054}}.

\bibitem{gunin1998relativistic}
A.~V. Gunin, A.~I. Klimov, S.~D. Korovin, I.~K. Kurkan, I.~V. Pegel, S.~D.
  Polevin, A.~M. Roitman, V.~V. Rostov, A.~S. Stepchenko, E.~M. Totmeninov,
  Relativistic {X}-band {BWO} with 3-{GW} output power, IEEE Trans. Plasma Sci.
  26 (1998) 326--331.
\newblock \href {http://dx.doi.org/10.1109/27.700761}
  {\path{doi:10.1109/27.700761}}.

\bibitem{case1984space}
W.~B. Case, R.~D. Kaplan, J.~E. Golub, J.~E. Walsh, Space-charge-{C}erenkov and
  cyclotron-{C}erenkov instabilities in an electron-beam dielectric system, J.
  Appl. Phys. 55 (1984) 2651--2658.
\newblock \href {http://dx.doi.org/http://dx.doi.org/10.1063/1.333275}
  {\path{doi:http://dx.doi.org/10.1063/1.333275}}.

\bibitem{chew1995waves}
W.~C. Chew, Waves and Fields in Inhomogeneous Media, IEEE press, New York,
  1995.

\bibitem{moreland1994efficiency}
L.~D. Moreland, E.~Schamiloglu, W.~Lemke, S.~D. Korovin, V.~V. Rostov, A.~M.
  Roitman, K.~J. Hendricks, T.~A. Spencer, Efficiency enhancement of high power
  vacuum {BWO}'s using nonuniform slow wave structures, IEEE Trans. Plasma Sci.
  22 (1994) 554--565.
\newblock \href {http://dx.doi.org/10.1109/27.338268}
  {\path{doi:10.1109/27.338268}}.

\bibitem{chipengo2015novel}
U.~Chipengo, M.~Zuboraj, N.~K. Nahar, J.~L. Volakis, A novel slow-wave
  structure for high-power-band backward wave oscillators with mode control,
  IEEE Trans. Plasma Sci. 43 (2015) 1879--1886.
\newblock \href {http://dx.doi.org/10.1109/TPS.2015.2431647}
  {\path{doi:10.1109/TPS.2015.2431647}}.

\bibitem{shiffler1991high}
D.~Shiffler, J.~A. Nation, L.~Schachter, J.~D. Ivers, G.~S. Kerslick, A
  high-power two stage traveling-wave tube amplifier, J. Appl. Phys. 70 (1991)
  106--113.
\newblock \href {http://dx.doi.org/http://dx.doi.org/10.1063/1.350322}
  {\path{doi:http://dx.doi.org/10.1063/1.350322}}.

\bibitem{agee1998evolution}
F.~J. Agee, Evolution of pulse shortening research in narrow band, high power
  microwave sources, IEEE Trans. Plasma Sci. 26 (1998) 235--245.
\newblock \href {http://dx.doi.org/10.1109/27.700749}
  {\path{doi:10.1109/27.700749}}.

\bibitem{korovin2000pulsewidth}
S.~D. Korovin, G.~A. Mesyats, I.~V. Pegel, S.~D. Polevin, V.~P. Tarakanov,
  Pulsewidth limitation in the relativistic backward wave oscillator, IEEE
  Trans. Plasma Sci. 28~(3) (2000) 485--495.
\newblock \href {http://dx.doi.org/10.1109/27.887654}
  {\path{doi:10.1109/27.887654}}.

\bibitem{vlasov2000overmoded}
A.~N. Vlasov, A.~G. Shkvarunets, J.~C. Rodgers, Y.~Carmel, T.~M. Antonsen,
  T.~M. Abuelfadl, D.~Lingze, V.~A. Cherepenin, G.~S. Nusinovich, M.~Botton,
  et~al., Overmoded {GW}-class surface-wave microwave oscillator, IEEE Trans.
  Plasma Sci. 28 (2000) 550--560.
\newblock \href {http://dx.doi.org/10.1109/PLASMA.2000.855071}
  {\path{doi:10.1109/PLASMA.2000.855071}}.

\bibitem{birdsall2004plasma}
C.~K. Birdsall, A.~B. Langdon, Plasma Physics via Computer Simulation, CRC
  Press, New York, 2004.

\bibitem{hockney1988computer}
R.~W. Hockney, J.~W. Eastwood, Computer Simulation Using Particles, CRC Press,
  New York, 1988.

\bibitem{grigoryev2002numerical}
Y.~N. Grigoryev, V.~A. Vshivkov, M.~P. Fedoruk, Numerical ``Particle-In-Cell"
  Methods: Theory and Applications, Walter de Gruyter, Boston, 2002.

\bibitem{candel2010parallel}
A.~Candel, A.~Kabel, L.~Lee, Z.~Li, C.~Ng, G.~Schussman, K.~Ko, I.~Ben-Zvi,
  J.~Kewisch, Parallel 3d finite element particle-in-cell simulations with
  pic3p, in: No. SLAC-PUB-13671, Vol.~2, 2009, p. FR5PFP069.

\bibitem{dawson1983particle}
J.~M. Dawson, Particle simulation of plasmas, Rev. Mod. Phys. 55 (1983)
  403--447.
\newblock \href {http://dx.doi.org/10.1103/RevModPhys.55.403}
  {\path{doi:10.1103/RevModPhys.55.403}}.

\bibitem{wang2009unipic}
J.~Wang, D.~Zhang, C.~Liu, Y.~Li, Y.~Wang, H.~Wang, H.~Qiao, X.~Li, Unipic code
  for simulations of high power microwave devices, Phys. Plasmas 16 (2009)
  033108.
\newblock \href {http://dx.doi.org/10.1063/1.3091931}
  {\path{doi:10.1063/1.3091931}}.

\bibitem{wang2010three}
J.~Wang, Z.~Chen, Y.~Wang, D.~Zhang, C.~Liu, Y.~Li, H.~Wang, H.~Qiao, M.~Fu,
  Y.~Yuan, Three-dimensional parallel unipic-3d code for simulations of
  high-power microwave devices, Phys. Plasmas 17 (2010) 073107.
\newblock \href {http://dx.doi.org/10.1063/1.3454766}
  {\path{doi:10.1063/1.3454766}}.

\bibitem{nieter2009application}
C.~Nieter, J.~R. Cary, G.~R. Werner, D.~N. Smithe, P.~H. Stoltz, Application of
  dey–mittra conformal boundary algorithm to 3d electromagnetic modeling, J.
  Comp. Phys. 228 (2009) 7902 -- 7916.
\newblock \href {http://dx.doi.org/http://dx.doi.org/10.1016/j.jcp.2009.07.025}
  {\path{doi:http://dx.doi.org/10.1016/j.jcp.2009.07.025}}.

\bibitem{taflove2000computational}
A.~Taflove, S.~C. Hagness, Computational Electrodynamics: The Finite-Difference
  Time-Domain Method, 3rd Edition, Artech House, Norwood, 2005.

\bibitem{greenwood2004elimination}
A.~D. Greenwood, K.~L. Cartwright, J.~W. Luginsland, E.~A. Baca, On the
  elimination of numerical {C}erenkov radiation in {PIC} simulations, J.
  Comput. Phys. 201 (2004) 665--684.
\newblock \href {http://dx.doi.org/http://dx.doi.org/10.1016/j.jcp.2004.06.021}
  {\path{doi:http://dx.doi.org/10.1016/j.jcp.2004.06.021}}.

\bibitem{meierbachtol2015conformal}
C.~S. Meierbachtol, A.~D. Greenwood, J.~P. Verboncoeur, B.~Shanker, Conformal
  electromagnetic particle in cell: A review, IEEE Trans. Plasma Sci. 43 (2015)
  3778--3793.
\newblock \href {http://dx.doi.org/10.1109/TPS.2015.2487522}
  {\path{doi:10.1109/TPS.2015.2487522}}.

\bibitem{wang2016conformal}
Y.~Wang, J.~Wang, Z.~Chen, G.~Cheng, P.~Wang, Three-dimensional simple
  conformal symplectic particle-in-cell methods for simulations of high power
  microwave devices, Comp. Phys. Comm. 205 (2016) 1 -- 12.
\newblock \href {http://dx.doi.org/http://dx.doi.org/10.1016/j.cpc.2016.03.007}
  {\path{doi:http://dx.doi.org/10.1016/j.cpc.2016.03.007}}.

\bibitem{teixeira2008time}
F.~L. Teixeira, Time-domain finite-difference and finite-element methods for
  {M}axwell equations in complex media, IEEE Trans. Antennas Propag. 56 (2008)
  2150--2166.
\newblock \href {http://dx.doi.org/10.1109/TAP.2008.926767}
  {\path{doi:10.1109/TAP.2008.926767}}.

\bibitem{lee1997time}
J.-F. Lee, R.~Lee, A.~Cangellaris, Time-domain finite-element methods, IEEE
  Trans. Antennas Propag. 45 (1997) 430--442.
\newblock \href {http://dx.doi.org/10.1109/8.558658}
  {\path{doi:10.1109/8.558658}}.

\bibitem{eastwood1991virtual}
J.~W. Eastwood, The virtual particle electromagnetic particle-mesh method,
  Comput. Phys. Commun. 64 (1991) 252--266.
\newblock \href
  {http://dx.doi.org/http://dx.doi.org/10.1016/0010-4655(91)90036-K}
  {\path{doi:http://dx.doi.org/10.1016/0010-4655(91)90036-K}}.

\bibitem{marder1987method}
B.~Marder, A method for incorporating {G}auss' law into electromagnetic {PIC}
  codes, J. Comput. Phys. 68 (1987) 48--55.
\newblock \href
  {http://dx.doi.org/http://dx.doi.org/10.1016/0021-9991(87)90043-X}
  {\path{doi:http://dx.doi.org/10.1016/0021-9991(87)90043-X}}.

\bibitem{burke1985}
W.~L. Burke, Applied Differential Geometry, Cambridge University Press,
  Cambridge, 1985.

\bibitem{flanders1989}
H.~Flanders, Differential Forms with Applications to the Physical Sciences,
  Dover, New York, 1989.

\bibitem{teixeira1999lattice}
F.~L. Teixeira, W.~C. Chew, Lattice electromagnetic theory from a topological
  viewpoint, J. Math. Phys. 40 (1999) 169--187.
\newblock \href {http://dx.doi.org/http://dx.doi.org/10.1063/1.532767}
  {\path{doi:http://dx.doi.org/10.1063/1.532767}}.

\bibitem{kotiuga2004}
P.~W. Gross, P.~R. Kotiuga, Electromagnetic Theory and Computation: A
  Topological Approach, Cambridge University Press, Cambridge, 2004.

\bibitem{he2007differential}
B.~He, F.~L. Teixeira, Differential forms, {G}alerkin duality, and sparse
  inverse approximations in finite element solutions of {M}axwell equations,
  IEEE Trans. Antennas Propag. 55 (2007) 1359--1368.
\newblock \href {http://dx.doi.org/10.1109/TAP.2007.895619}
  {\path{doi:10.1109/TAP.2007.895619}}.

\bibitem{teixeira2014lattice}
F.~L. Teixeira, Lattice {M}axwell's equations, Prog. Electromagn. Res. 148
  (2014) 113--128.
\newblock \href {http://dx.doi.org/10.2528/PIER14062904}
  {\path{doi:10.2528/PIER14062904}}.

\bibitem{moon2015exact}
H.~Moon, F.~L. Teixeira, Y.~A. Omelchenko, Exact charge-conserving
  scatter–gather algorithm for particle-in-cell simulations on unstructured
  grids: A geometric perspective, Comput. Phys. Commun. 194 (2015) 43--53.
\newblock \href {http://dx.doi.org/http://dx.doi.org/10.1016/j.cpc.2015.04.014}
  {\path{doi:http://dx.doi.org/10.1016/j.cpc.2015.04.014}}.

\bibitem{squire2012geometric}
J.~Squire, H.~Qin, W.~M. Tang, Geometric integration of the {V}lasov-{M}axwell
  system with a variational particle-in-cell scheme, Phys. Plasmas 19 (2012)
  084501.
\newblock \href {http://dx.doi.org/http://dx.doi.org/10.1063/1.4742985}
  {\path{doi:http://dx.doi.org/10.1063/1.4742985}}.

\bibitem{pinto2014charge}
M.~C. Pinto, S.~Jund, S.~Salmon, E.~Sonnendrucker, Charge-conserving
  {FEM}-{PIC} schemes on general grids, C. R. Mec. 342 (2014) 570--582.
\newblock \href
  {http://dx.doi.org/http://dx.doi.org/10.1016/j.crme.2014.06.011}
  {\path{doi:http://dx.doi.org/10.1016/j.crme.2014.06.011}}.

\bibitem{pinto2016divergence}
M.~C. Pinto, M.~Mounier, E.~Sonnendrucker, Handling the divergence constraints
  in {M}axwell and {V}lasov-{M}axwell simulations, Appl. Math. Comp. 272 (2016)
  403--419.
\newblock \href {http://dx.doi.org/http://dx.doi.org/10.1016/j.amc.2015.07.089}
  {\path{doi:http://dx.doi.org/10.1016/j.amc.2015.07.089}}.

\bibitem{he2006sparse}
B.~He, F.~L. Teixeira, Sparse and explicit {FETD} via approximate inverse
  {H}odge (mass) matrix, IEEE Microw. Wireless Compon. Lett. 16 (2006)
  348--350.

\bibitem{kim2011parallel}
J.~Kim, F.~L. Teixeira, Parallel and explicit finite-element time-domain method
  for {M}axwell's equations, IEEE Trans. Antennas Propag. 59 (2011) 2350--2356.
\newblock \href {http://dx.doi.org/10.1109/TAP.2011.2143682}
  {\path{doi:10.1109/TAP.2011.2143682}}.

\bibitem{na2016local}
D.-Y. Na, H.~Moon, Y.~A. Omelchenko, F.~L. Teixeira, Local, explicit, and
  charge-conserving electromagnetic particle-in-cell algorithm on unstructured
  grids, IEEE Trans. Plasma Sci. 44 (2016) 1353--1362.
\newblock \href {http://dx.doi.org/10.1109/TPS.2016.2582143}
  {\path{doi:10.1109/TPS.2016.2582143}}.

\bibitem{teixeira1999differential}
F.~Teixeira, W.~Chew, Differential forms, metrics, and the reflectionless
  absorption of electromagnetic waves, J. Electromagn. Waves Appl. 13 (1999)
  665--686.
\newblock \href {http://dx.doi.org/http://dx.doi.org/10.1163/156939399X01104}
  {\path{doi:http://dx.doi.org/10.1163/156939399X01104}}.

\bibitem{pendry2006controlling}
J.~B. Pendry, D.~Schurig, D.~R. Smith, Controlling electromagnetic fields,
  Science 312 (2006) 1780--1782.
\newblock \href {http://dx.doi.org/10.1126/science.1125907}
  {\path{doi:10.1126/science.1125907}}.

\bibitem{chen2010nature}
H.~Chen, C.~T. Chan, P.~Sheng, Transformation optics and metamaterials, Nature
  Mater. 9 (2010) 387--396.
\newblock \href {http://dx.doi.org/10.1038/nmat2743}
  {\path{doi:10.1038/nmat2743}}.

\bibitem{verboncoeur2005particle}
J.~P. Verboncoeur, Particle simulation of plasmas: {R}eview and advances,
  Plasma Phys. and Contr. F. 47 (2005) A231--A260.
\newblock \href {http://dx.doi.org/doi:10.1088/0741-3335/47/5A/017}
  {\path{doi:doi:10.1088/0741-3335/47/5A/017}}.

\bibitem{vay2008simulation}
J.-L. Vay, Simulation of beams or plasmas crossing at relativistic velocity,
  Phys. Plasmas 15 (2008) 056701.
\newblock \href {http://dx.doi.org/http://dx.doi.org/10.1063/1.2837054}
  {\path{doi:http://dx.doi.org/10.1063/1.2837054}}.

\bibitem{FEMSTER2005}
P.~Castillo, R.~Rieben, D.~White, {F}{E}{M}{S}{T}{E}{R}: An object-oriented
  class library of high-order discrete differential forms, ACM Trans. Math.
  Software 31~(4) (2005) 425 -- 457.

\bibitem{warnick1997teaching}
K.~F. Warnick, R.~H. Selfridge, D.~V. Arnold, Teaching electromagnetic field
  theory using differential forms, IEEE Trans. Edu. 40 (1997) 53--68.
\newblock \href {http://dx.doi.org/10.1109/13.554670}
  {\path{doi:10.1109/13.554670}}.

\bibitem{sen2000geometric}
S.~Sen, S.~Sen, J.~C. Sexton, D.~H. Adams, Geometric discretization scheme
  applied to the abelian chern-simons theory, Phys. Rev. E 61 (2000)
  3174--3185.
\newblock \href {http://dx.doi.org/10.1103/PhysRevE.61.3174}
  {\path{doi:10.1103/PhysRevE.61.3174}}.

\bibitem{bossavit1988whitney}
A.~Bossavit, Whitney forms: A class of finite elements for three-dimensional
  computations in electromagnetism, IEE Proc., Part A: Phys. Sci., Meas.
  Instrum., Manage. Educ. 135 (1988) 493--500.
\newblock \href {http://dx.doi.org/10.1049/ip-a-1.1988.0077}
  {\path{doi:10.1049/ip-a-1.1988.0077}}.

\bibitem{teixeira2013differential}
F.~L. Teixeira, Differential forms in lattice field theories: An overview, ISRN
  Math. Phys. 2013 (2013) 16.
\newblock \href {http://dx.doi.org/http://dx.doi.org/10.1155/2013/487270}
  {\path{doi:http://dx.doi.org/10.1155/2013/487270}}.

\bibitem{tarhasaari1999some}
T.~Tarhasaari, L.~Kettunen, A.~Bossavit, Some realizations of a discrete
  {H}odge operator: a reinterpretation of finite element techniques [for {EM}
  field analysis], IEEE Trans. Magn. 35 (1999) 1494--1497.
\newblock \href {http://dx.doi.org/10.1109/20.767250}
  {\path{doi:10.1109/20.767250}}.

\bibitem{gillette2011dual}
A.~Gillette, C.~Bajaj, Dual formulations of mixed finite element methods with
  applications, Comput. Aided Des. 43 (2011) 1213--1221.
\newblock \href {http://dx.doi.org/http://dx.doi.org/10.1016/j.cad.2011.06.017}
  {\path{doi:http://dx.doi.org/10.1016/j.cad.2011.06.017}}.

\bibitem{white2015mixed}
R.~Rieben, G.~Rodrigue, D.~White, A high order mixed vector finite element
  method for solving the time dependent {M}axwell equations on unstructured
  grids, J. Comp. Phys. 204~(2) (2005) 490 -- 519.
\newblock \href {http://dx.doi.org/http://dx.doi.org/10.1016/j.jcp.2004.10.030}
  {\path{doi:http://dx.doi.org/10.1016/j.jcp.2004.10.030}}.

\bibitem{he2005degrees}
B.~He, F.~L. Teixeira, On the degrees of freedom of lattice electrodynamics,
  Phys. Lett. A 336 (2005) 1--7.
\newblock \href
  {http://dx.doi.org/http://dx.doi.org/10.1016/j.physleta.2005.01.001}
  {\path{doi:http://dx.doi.org/10.1016/j.physleta.2005.01.001}}.

\bibitem{hughes1981lagrangian}
T.~J. Hughes, W.~K. Liu, T.~K. Zimmermann, Lagrangian-{E}ulerian finite element
  formulation for incompressible viscous flows, Comput. Method Appl. M. 29
  (1981) 329--349.
\newblock \href
  {http://dx.doi.org/http://dx.doi.org/10.1016/0045-7825(81)90049-9}
  {\path{doi:http://dx.doi.org/10.1016/0045-7825(81)90049-9}}.

\bibitem{guth1980existence}
A.~H. Guth, Existence proof of a nonconfining phase in four-dimensional {U}(1)
  lattice gauge theory, Phys. Rev. D 21 (1980) 2291--2307.
\newblock \href {http://dx.doi.org/10.1103/PhysRevD.21.2291}
  {\path{doi:10.1103/PhysRevD.21.2291}}.

\bibitem{he2006geometric}
B.~He, F.~L. Teixeira, Geometric finite element discretization of {M}axwell
  equations in primal and dual spaces, Phys. Lett. A 349 (2006) 1--14.
\newblock \href
  {http://dx.doi.org/http://dx.doi.org/10.1016/j.physleta.2005.09.002}
  {\path{doi:http://dx.doi.org/10.1016/j.physleta.2005.09.002}}.

\bibitem{donderici2008mixed}
B.~Donderici, F.~L. Teixeira, Mixed finite-element time-domain method for
  transient {M}axwell equations in doubly dispersive media, IEEE Trans. Microw.
  Theory Techn. 56 (2008) 113--120.
\newblock \href {http://dx.doi.org/10.1109/TMTT.2007.912217}
  {\path{doi:10.1109/TMTT.2007.912217}}.

\bibitem{teixeira1998general}
F.~L. Teixeira, W.~C. Chew, General closed-form pml constitutive tensors to
  match arbitrary bianisotropic and dispersive linear media, IEEE Microw.
  Guided Wave Lett. 8 (1998) 223--225.
\newblock \href {http://dx.doi.org/10.1109/75.678571}
  {\path{doi:10.1109/75.678571}}.

\bibitem{teixeira1999unified}
F.~L. Teixeira, W.~C. Chew, Unified analysis of perfectly matched layers using
  differential forms, Microw. Opt. Techn. Lett. 20 (1999) 124--126.
\newblock \href
  {http://dx.doi.org/10.1002/(SICI)1098-2760(19990120)20:2<124::AID-MOP12>3.0.CO;2-N}
  {\path{doi:10.1002/(SICI)1098-2760(19990120)20:2<124::AID-MOP12>3.0.CO;2-N}}.

\bibitem{WangTPS2006}
J.~Wang, Y.~Wang, D.~Zhang, Truncation of open boundaries of cylindrical
  waveguides in 2.5-dimensional problems by using the convolutional perfectly
  matched layer, IEEE Trans. Plasma Sci. 34 (2006) 681.

\bibitem{kostov2002space}
K.~G. Kostov, J.~J. Barroso, Space-charge-limited current in cylindrical diodes
  with finite-length emitter, Phys. Plasmas 9 (2002) 1039--1042.
\newblock \href {http://dx.doi.org/10.1063/1.1446876}
  {\path{doi:10.1063/1.1446876}}.

\bibitem{benford2015high}
J.~Benford, J.~A. Swegle, E.~Schamiloglu, High Power Microwaves, CRC Press, New
  York, 2015.

\bibitem{barroso2002cylindrical}
J.~J. Barroso, J.~P.~L. Neto, K.~G. Kostov, Cylindrical waveguide with axially
  rippled wall, J. Microw. Optoelectron. Electromagn. Appl. 2 (2002) 75--89.

\bibitem{cairns1936the}
S.~S. Cairns, The generalized theorem of {S}tokes, Trans. Amer. Math. Soc. 40
  (1936) 167--174.

\end{thebibliography}

\end{document}